\documentclass[twocolumn]{aastex63}
\usepackage{amsmath}
\usepackage{lineno}

\def\ltsima{$\; \buildrel < \over \sim \;$}
\def\simlt{\lower.5ex\hbox{\ltsima}}
\def\gtsima{$\; \buildrel > \over \sim \;$}
\def\simgt{\lower.5ex\hbox{\gtsima}}
\def\kms{{\rm\,km\,s^{-1}}}
\def\mas{{\rm\,mas}}
\def\masyr{{\rm\,mas/yr}}
\def\kpc{{\rm\,kpc}}

\def\pc{{\rm\,pc}}

\def\g{{\rm\,g}}

\def \r {r$^{1/4}$ }

\newcommand{\fmmm}[1]{\mbox{$#1$}}
\newcommand{\scnd}{\mbox{\fmmm{''}\hskip-0.3em .}}


\def\AA{$\; \buildrel \circ \over {\rm A}$}

\def\deg{^\circ}

\def\s{\ifmmode \widetilde \else \~\fi}
\def\={\overline}

\def\spose#1{\hbox to 0pt{#1\hss}}

\def\lta{\mathrel{\spose{\lower 3pt\hbox{$\mathchar"218$}}
     \raise 2.0pt\hbox{$\mathchar"13C$}}}
\def\gta{\mathrel{\spose{\lower 3pt\hbox{$\mathchar"218$}}
     \raise 2.0pt\hbox{$\mathchar"13E$}}}
\def\Dt{\spose{\raise 1.5ex\hbox{\hskip3pt$\mathchar"201$}}}    
\def\dt{\spose{\raise 1.0ex\hbox{\hskip2pt$\mathchar"201$}}}    

\def\r{${\rm r^{1/4}}$}

\def\dotsfill{\leaders\hbox to 1em{\hss.\hss}\hfill}
\def\sun{\odot}
\def\earth{\oplus}

\def\Gyr{{\rm\,Gyr}}

\def\ltsima{$\; \buildrel < \over \sim \;$}
\def\gtsima{$\; \buildrel > \over \sim \;$}
\def\lsim{\lower.5ex\hbox{\ltsima}}
\def\gsim{\lower.5ex\hbox{\gtsima}}
\def\lapp{\ifmmode\stackrel{<}{_{\sim}}\else$\stackrel{<}{_{\sim}}$\fi}
\def\gapp{\ifmmode\stackrel{>}{_{\sim}}\else$\stackrel{<}{_{\sim}}$\fi}

\submitjournal{ApJ}

\shorttitle{A search for stellar streams with Gaia DR2 and EDR3}
\shortauthors{Ibata et al.}
\graphicspath{{./}{figures/}}

\begin{document}

\title{Charting the Galactic acceleration field \\
I. A search for stellar streams with Gaia DR2 and EDR3\\
with follow-up from ESPaDOnS and UVES}

\correspondingauthor{Rodrigo Ibata}
\email{rodrigo.ibata@astro.unistra.fr}

\author[0000-0002-3292-9709]{Rodrigo Ibata}
\affiliation{Universit\'e de Strasbourg, CNRS, Observatoire astronomique de Strasbourg, UMR 7550, F-67000 Strasbourg, France}
\nocollaboration{1}

\author[0000-0002-8318-433X]{Khyati Malhan}
\affiliation{The Oskar Klein Centre, Department of Physics, Stockholm University, AlbaNova, SE-10691 Stockholm, Sweden}
\nocollaboration{1}

\author[0000-0002-1349-202X]{Nicolas Martin}
\affiliation{Universit\'e de Strasbourg, CNRS, Observatoire astronomique de Strasbourg, UMR 7550, F-67000 Strasbourg, France}
\affiliation{Max-Planck-Institut f\"ur Astronomie, K\"onigstuhl 17, D-69117, Heidelberg, Germany}
\nocollaboration{1}

\author{Dominique Aubert}
\affiliation{Universit\'e de Strasbourg, CNRS, Observatoire astronomique de Strasbourg, UMR 7550, F-67000 Strasbourg, France}
\nocollaboration{1}

\author[0000-0003-3180-9825]{Benoit Famaey}
\affiliation{Universit\'e de Strasbourg, CNRS, Observatoire astronomique de Strasbourg, UMR 7550, F-67000 Strasbourg, France}
\nocollaboration{1}

\author[0000-0002-0358-4502]{Paolo Bianchini}
\affiliation{Universit\'e de Strasbourg, CNRS, Observatoire astronomique de Strasbourg, UMR 7550, F-67000 Strasbourg, France}
\nocollaboration{1}

\author[0000-0002-6863-0661]{Giacomo Monari}
\affiliation{Universit\'e de Strasbourg, CNRS, Observatoire astronomique de Strasbourg, UMR 7550, F-67000 Strasbourg, France}
\nocollaboration{1}

\author[0000-0001-8059-2840]{Arnaud Siebert}
\affiliation{Universit\'e de Strasbourg, CNRS, Observatoire astronomique de Strasbourg, UMR 7550, F-67000 Strasbourg, France}
\nocollaboration{1}

\author[0000-0002-2468-5521]{Guillaume F. Thomas}
\affiliation{Instituto de Astrof\'isica de Canarias, E-38205 La Laguna, Tenerife, Spain}
\affiliation{Universidad de La Laguna, Dpto. Astrof\'isica, E-38206 La Laguna, Tenerife, Spain}
\nocollaboration{1}

\author[0000-0001-8200-810X]{Michele Bellazzini}
\affiliation{INAF - Osservatorio di Astrofisica e Scienza dello Spazio, via Gobetti 93/3, I-40129 Bologna, Italy}
\nocollaboration{1}

\author{Piercarlo Bonifacio}
\affiliation{GEPI, Observatoire de Paris, Universit\'e PSL, CNRS, 5 Place Jules Janssen, 92190 Meudon, France}
\nocollaboration{1}

\author{Elisabetta Caffau}
\affiliation{GEPI, Observatoire de Paris, Universit\'e PSL, CNRS, 5 Place Jules Janssen, 92190 Meudon, France}
\nocollaboration{1}

\author[0000-0001-5073-2267]{Florent Renaud}
\affiliation{Department of Astronomy and Theoretical Physics, Lund Observatory, Box 43, 221 00 Lund, Sweden}
\nocollaboration{1}

\begin{abstract}
We present maps of the stellar streams detected in the {\it Gaia} Data Release 2 (DR2) and Early Data Release 3 (EDR3) catalogs using the \texttt{STREAMFINDER} algorithm. We also report the spectroscopic follow-up of the brighter DR2 stream members obtained with the high-resolution CFHT/ESPaDOnS and VLT/UVES spectrographs as well as with the medium-resolution NTT/EFOSC2 spectrograph. Two new stellar streams that do not have a clear progenitor are detected in DR2 (named Hr\'{\i}d and Gunnthr\'a), and seven are detected in EDR3 (named Gaia-6 to Gaia-12). Several candidate streams are also identified. The software also finds very long tidal tails associated with the 15 globular clusters NGC 288, NGC~1261, NGC 1851,  NGC~2298, NGC~2808, NGC~3201, M~68, $\omega$Cen, NGC 5466, Palomar~5, M~5, NGC~6101, M~92, NGC~6397 and NGC~7089. These stellar streams will be used in subsequent contributions in this series to chart the properties of the Galactic acceleration field on $\sim 100\pc$ to $\sim 100\kpc$ scales.
\end{abstract}


\section{Introduction}
\label{sec:Introduction}

The distribution of matter in our Galaxy and the fundamental behavior of the gravitational force are encoded in the three-dimensional acceleration field of the Milky Way. If we were able to measure the acceleration field accurately, we could assess whether the dark matter distribution is consistent with predictions from standard Lambda Cold Dark Matter cosmology \citep{2008MNRAS.391.1685S}, or whether alternative dark matter models (warm, fuzzy, self-interacting, etc) explain the acceleration field more naturally. Given the absence of a direct detection of a dark matter particle, despite decades of searches \citep[e.g.,][]{2019JPhG...46j3003S}, it is also conceivable that the gravitational force does not behave as predicted by General Relativity at galactic scales \citep[e.g.,][]{2020arXiv200700082S}. Such a  possibility may also be explored with an accurate map of the acceleration field throughout our Galaxy.

It was with these goals in mind that our team embarked on a search for stellar streams in the Milky Way. Dynamically cold stellar streams offer an opportunity to probe the acceleration field locally over the extent of the stream structure \citep{2002MNRAS.332..915I, 2002ApJ...570..656J, 2012ApJ...748...20C}, and also globally in the host galaxy throughout the spatial volume that the progenitor satellite travelled through \citep{1999ApJ...512L.109J}. Such streams are formed from stars lost to dissolving globular clusters (and perhaps very low mass dwarf satellite galaxies) as they orbit around their host galaxy. Internal or external processes (or a combination of both) can drive the dissolution. Among the internal processes that dissolve clusters are evaporation via two-body scatterings and collectively amplified fluctuations, or ejection from three-body interactions with binary stars \citep[e.g.,][]{2008gady.book.....B}. External processes which shorten the clusters lifetime include tidal disruption, disk shocking and interactions with other substructures, such as giant molecular clouds, spiral arms, the bar, dark matter substructures, etc \citep{Amorisco2016, Hattori2016}.

In situations where the progenitors are of low mass and dissolve slowly, and where the ejected stars are lost with low relative energy, the resulting tidal streams follow closely the progenitor's orbit. When the progenitors have larger mass, the ejected stars have to overcome the internal potential well in their journey out from the satellite, and as a consequence they can emerge with a significantly different velocity to the progenitor \citep{2008MNRAS.387.1248K}. While they thus no longer follow the same orbit, the new path (and hence the locus of the observed tidal stream structure) can be readily calculated with N-body simulations (or with the Lagrange-point or ``streak-line'' approximation, \citealt{2011MNRAS.417..198V,2012MNRAS.420.2700K}). This procedure can be inverted: N-body simulations can be fit to an observational configuration to deduce the orbit of the progenitor. The orbit, of course, gives direct access to the underlying acceleration field. This is the approach we intend to implement in this series of papers.

The observational foundation of this study is a survey of stream-like structures detected in the {\it Gaia} Data Release 2 (DR2) \citep{2018A&A...616A...2L, 2018A&A...616A...1G} and Early Data Release 3 (EDR3)  \citep{GaiaEDR3_Brown_2020, GaiaEDR3_Lindegren_2020, GaiaEDR3_Riello_2020, GaiaDR2_2018_Brown} catalogs using the \texttt{STREAMFINDER} algorithm \citep{2018MNRAS.477.4063M}. Although a number of these streams were known before {\it Gaia}, 13 were detected with the \texttt{STREAMFINDER} in Gaia \citep{2018MNRAS.481.3442M,2019ApJ...872..152I}, and we will report several more new detections and candidate detections below. Over the last two and a half years the brightest stars in these streams were targeted for spectroscopic follow-up with the high-resolution CFHT/ESPaDOnS and VLT/UVES instruments, with some preliminary spectra obtained at medium resolution with NTT/EFOSC. The present work is enabled by the combination of the excellent {\it Gaia} astrometry and the excellent radial velocity measurements from ESPaDOnS and UVES.

Our aim in the present study is first to update the {\it Gaia} DR2 stream maps shown previously in \citet[][hereafter Paper~I]{2019ApJ...872..152I} (Section~\ref{sec:DR2_maps}), and from which the targets of the spectroscopic campaign were drawn. We summarize the ESPaDOnS and UVES observational campaigns in Section~\ref{sec:Observations}. The detections made with DR2 data (and followed-up with spectroscopy) are presented in Section~\ref{sec:DR2_Detections}. Deeper maps based on the recently published {\it Gaia} EDR3 catalog are presented in Section~\ref{sec:EDR3_maps}, and the new findings with the EDR3 catalog are discussed in Section~\ref{sec:EDR3_Detections}. Finally, in Section~\ref{sec:Conclusions} we discuss the findings and draw the conclusions of this study.

Throughout the remainder of this study, when referring to the proper motion components in some spherical coordinate system (say $\alpha, \delta$): $\mu^*_\alpha (\equiv \mu_\alpha \cos(\delta))$, $\mu_\delta$, we will for convenience drop the asterix superscript from $\mu^*_\alpha$. All the extinction-corrected magnitudes listed in this work were obtained by assuming that the interstellar extinction is in the foreground, and interpolating the reddening value using the \citet{1998ApJ...500..525S} maps, as re-calibrated by \citet{2011ApJ...737..103S} with $R_V=3.1$.

\begin{figure*}
\begin{center}
{\Large Gaia DR2 detections, $[1,30]\kpc$}\par
\includegraphics[angle=0, viewport= 45 45 657 650, clip, width=16cm]{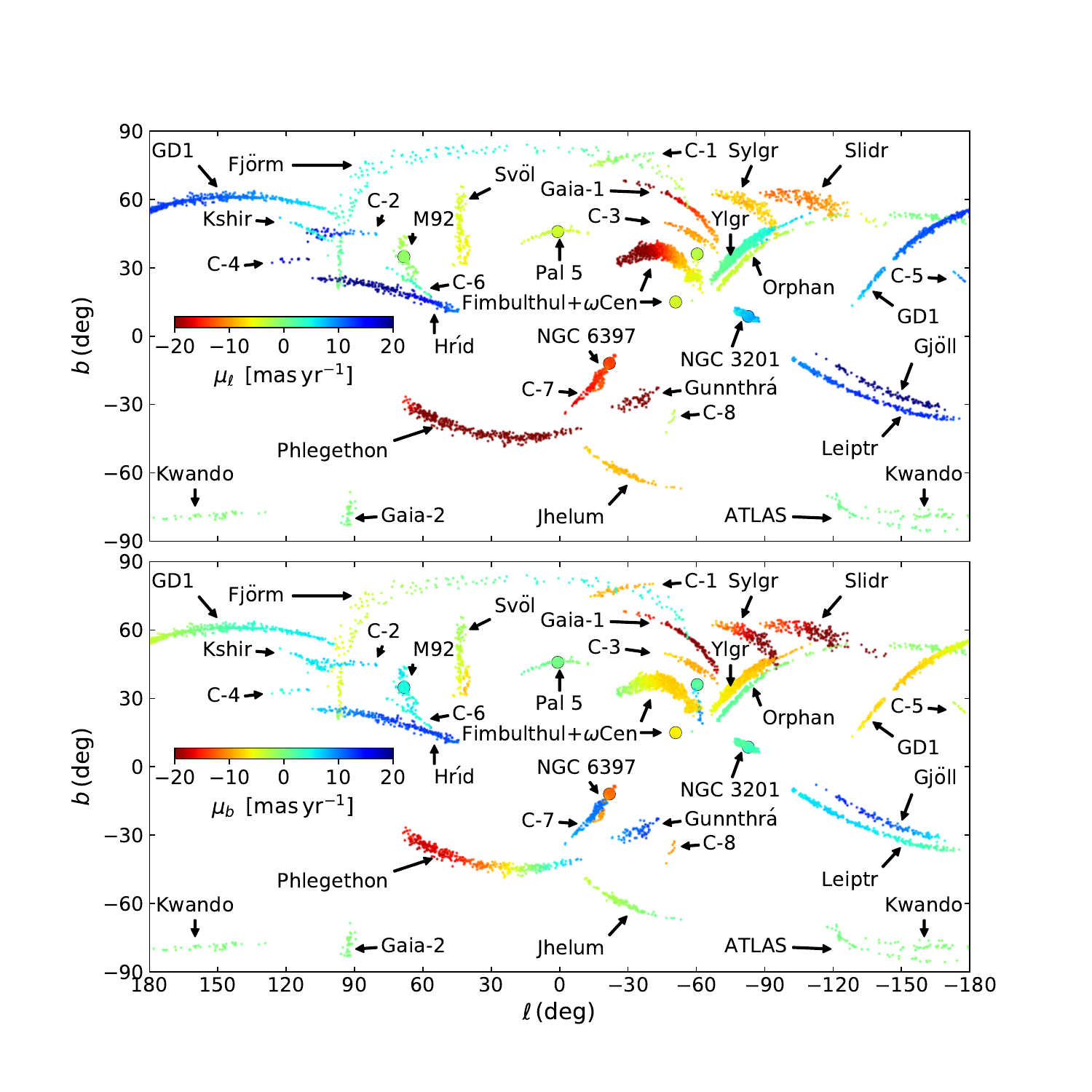}
\end{center}
\caption{Proper motion distribution in the Galactic longitude direction (top) and the Galactic latitude direction (bottom) of the 5960 stars in the sample derived from the {\it Gaia} DR2 catalog. These sources have a $>7\sigma$ likelihood of being stream members according to the {\tt STREAMFINDER} algorithm, and were used as the basis of the spectroscopic follow-up campaign. The stellar streams are identified by name, and include two new features labelled Hr\'{\i}d and Gunnthr\'a. A further 8 candidate streams (marked C-1 to C-8) were selected for observation, but additional spectroscopic information is still required for their confirmation. The six globular clusters NGC~3201, M~68 (not labelled, but at $\ell=-60\deg,b=36\deg$), $\omega$Cen, Palomar~5, M~92 and NGC~6397 create tidal tails that are detected in these maps.}
\label{fig:PMdist_all}
\end{figure*}

\section{Gaia DR2 \texttt{STREAMFINDER} maps}
\label{sec:DR2_maps}

The \texttt{STREAMFINDER} algorithm was designed to hunt for stream-like over-densities in photometric and astrometric catalogs. This would be a relatively trivial task if the stellar distances and radial velocities were measured in addition to sky position and proper motion. One would simply calculate the adiabatic-invariant actions and group together stars that are close to each other in action space. Although {\it Gaia} provides superb proper motion measurements even for faint stars, the parallax measurements are relatively poor for distant populations, and radial velocities are only available for the very brightest sources (at this time, to $G \sim 12.5$~mag). The missing distance and velocity information make it challenging to calculate the action variables.

A full description of the \texttt{STREAMFINDER} method is given in \citet{2018MNRAS.477.4063M} and Paper~I. Briefly, the algorithm circumvents the missing distance data by testing a series of stellar populations templates, and it circumvents the missing velocity data by scanning through heliocentric radial velocity. The \texttt{STREAMFINDER} is essentially a friend-finding algorithm that examines, in turn, every star in the input dataset. The star's trial orbit is calculated given the measured proper motion, the measured position, the estimated distance (consistent with the measured photometry and the trial isochrone model), for each radial velocity value sampled. (The measured radial velocity is used for the small subset of stars for which the radial velocity is actually known.) 

\begin{figure*}
\begin{center}
{\Large Gaia DR2 detections, $[1,30]\kpc$}\par
\includegraphics[angle=0, viewport= 45 45 657 650, clip, width=16cm]{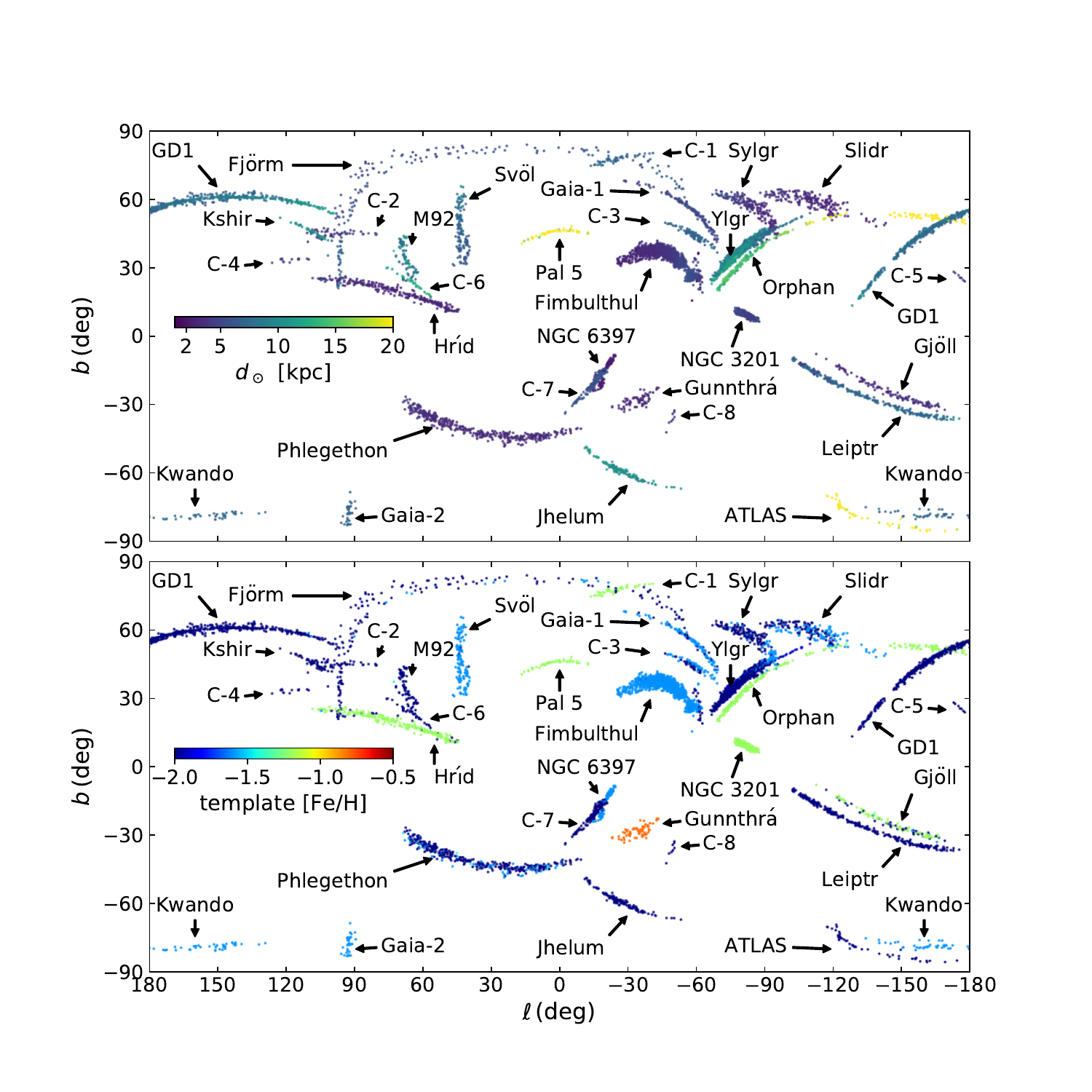}
\end{center}
\caption{As Figure~\ref{fig:PMdist_all} (i.e. for Gaia DR2), but showing the heliocentric distance solutions (top) found by the {\tt STREAMFINDER} software, and the corresponding metallicity of the stellar populations template (bottom).}
\label{fig:dist_and_FeH}
\end{figure*}

We construct a model of the probability density function of the stream $\mathcal{P}_{\rm stream}$, which is simply a smeared out version of the trial orbit of length $L$ (centered on the star), with Gaussian dispersion in sky position, distance, proper motion and radial velocity. As in Paper~I, we adopted $L=20\deg$, the stream model width was chosen to be $w_s=50\pc$ (we adopt the convention of using $w_x$ to denote model dispersion and $\sigma_x$ to denote measurement uncertainty in a variable $x$), and we took a stream velocity dispersion of $w_v = 3\kms$ in radial velocity, together with an identical dispersion in proper motion $w_\mu$ (given the estimated distance $d$). We consider this stream model to be present in addition to a ``contamination'' $\mathcal{P}_{\rm cont}$ due to the normal smooth population of the Galaxy. The logarithm of the likelihood of this stream model is thus:
\begin{equation}
{\ln} \mathcal{L} = \rm \sum_{\rm{data}} {\ln} \, [\eta \mathcal{P}_{\rm stream}(\theta) + (1 - \eta) \, \mathcal{P}_{\rm cont} ] \, ,
\label{eqn:likelihood}
\end{equation}
where $\theta$ represents the stream parameters and $\eta$ is the fraction of stars in the structure. For each trial distance obtained with the adopted single stellar population (SSP) model, the algorithm samples over the missing velocity information to find the highest value of $\mathcal{L}$ and the corresponding value of $\eta$. This allows us to answer the question: if there were a stream passing through the (phase-space) location of the star under consideration, how likely is the most likely stream model, and what fraction of stars are part of the structure?

\begin{figure*}
\begin{center}
{\Large Gaia DR2 detections, $[1,30]\kpc$}\par
\includegraphics[angle=0, viewport= 45 45 657 650, clip, width=16cm]{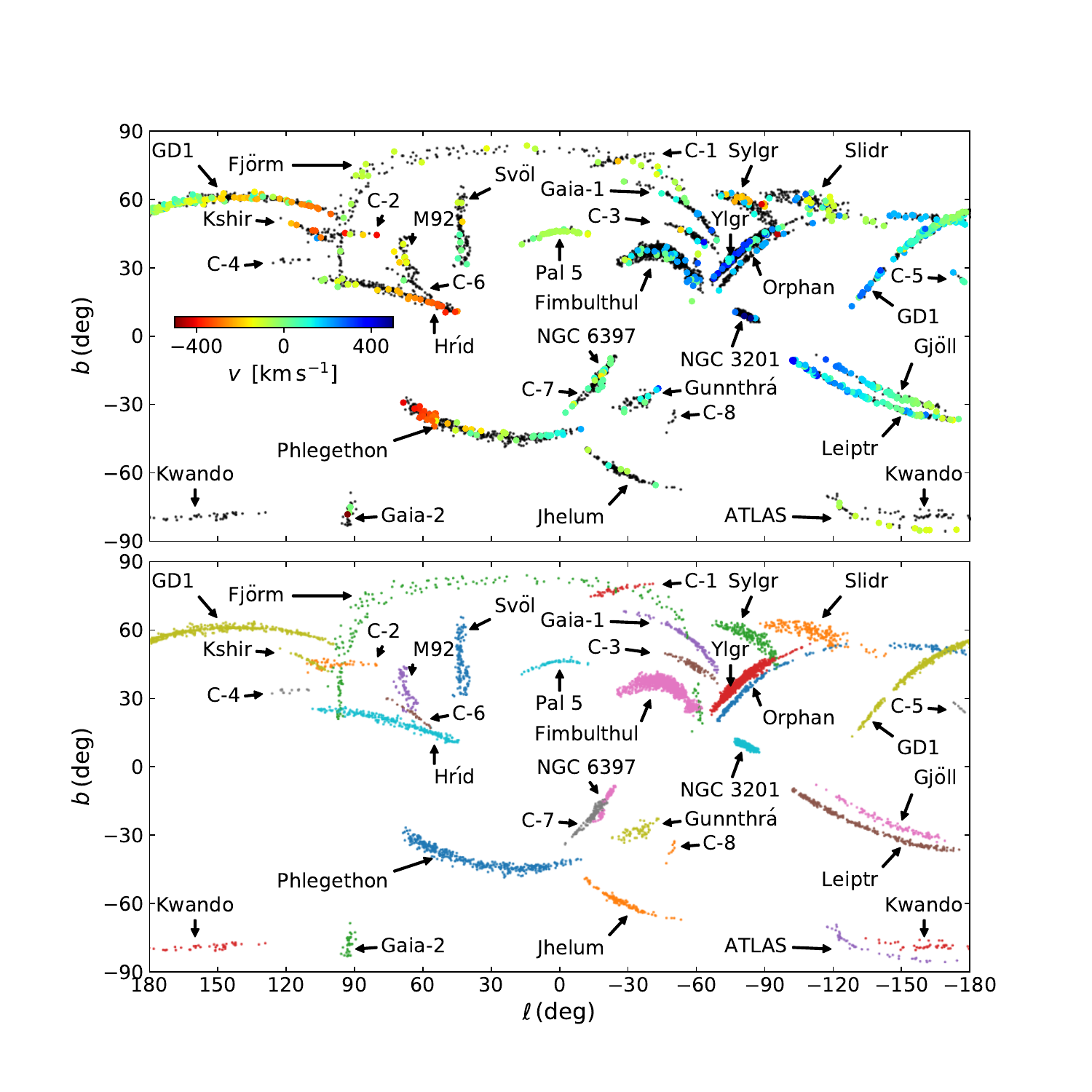}
\end{center}
\caption{As Figure~\ref{fig:PMdist_all}, but showing the sky distribution of the 685 radial velocity measurements of the DR2 sample (top). All of the streams show kinematic coherence, but with varying levels of contamination, dependent on stream distance and the local contrast over the ``normal'' populations of the Milky Way. In the bottom panel the different streams have been color-coded to allow for easier visual discrimination in sky regions where they overlap.}
\label{fig:map_vels}
\end{figure*}

The precise form of $\mathcal{P}_{\rm stream}$ is explained in Paper I, but briefly it is:
\begin{equation}
\mathcal{P}_{\rm stream}(\theta)  = \mathcal{P}_{\rm length} \times \mathcal{P}_{\rm width} \times \mathcal{P}_{\rm LF} \times \mathcal{P}_{\rm color} \times \mathcal{P}_\mu \times \mathcal{P}_{\rm \varpi}
\label{eqn:prob}
\end{equation}
where $\mathcal{P}_{\rm length}=1/L$ is the uniform probability in position along the stream, while $\mathcal{P}_{\rm width}= \mathcal{N}(\Delta s, w_s)$ is the Gaussian probability of being at a perpendicular distance $\Delta s$ away from the stream track. $\mathcal{P}_{\rm LF}$  is the probability that the star is drawn from the luminosity function of the adopted SSP model. $P_{\rm color} =  \mathcal{N}(\Delta (G_{\rm Bp} - G_{\rm Rp}), \sigma_{G_{\rm Bp} - G_{\rm Rp}})$ is the Gaussian probability of the measured color offset $\Delta (G_{\rm Bp} - G_{\rm Rp})$ from the adopted isochrone given the measured color uncertainty $\sigma_{G_{\rm Bp} - G_{\rm Rp}}$. The proper motion probability term $\mathcal{P}_\mu$ is a two-dimensional Gaussian PDF incorporating the proper motion uncertainties in right ascension and declination and their cross-terms (as explained in Paper I). 

In contrast to the stream PDF model in Paper~I, we now include the term $\mathcal{P}_{\rm \varpi}$ in Equation~\ref{eqn:prob}, which accounts for the Gaussian probability of the (inverse) trial distance to the star given the observed parallax $\varpi$ and its uncertainty. The small $0.029$~mas parallax bias of {\it Gaia} DR2 \citep{2018A&A...616A...2L} is added to the stream model at this stage. 

This new inclusion of parallax in Equation~\ref{eqn:prob} means that we cannot simply reuse the empirical contamination model $\mathcal{P}_{\rm cont}$ of Paper~I. We re-create it in essentially the same way as we did in that contribution, but with an additional parallax dimension. Briefly, we generate 1000 realizations of the {\it Gaia} catalog in which the stellar coordinates are each shifted with a $2\deg$ Gaussian random deviate. We combine these smoothed catalogs into sky maps (with pixels of $1.4\deg\times1.4\deg$ in zenithal equal area projection) as a function of color and magnitude (using intervals of $0.05 \, {\rm mag}$ and $0.10 \, {\rm mag}$, respectively). We refer to this as the fine spatial map $P_{\rm fine}(\alpha,\delta,G,G_{\rm BP}-G_{\rm RP})$, which is identical to that of Paper~I. In a similar way to that contribution, we use the {\tt Armadillo} software package \citep{Sanderson2016} to create Gaussian Mixture Model (GMM) decompositions in larger $5.6\deg\times5.6\deg$ spatial bins to the color-magnitude, the proper motion, and (now new) the parallax information. We fit this five-dimensional model $P_{\rm GMM}(\mu_\alpha,\mu_\delta,G,G_{\rm BP}-G_{\rm RP},\varpi | \alpha,\delta)$ using 100 GMM components, including cross-terms (the same number as used previously in Paper~I). Cutting through this model at the observed $G,G_{\rm BP}-G_{\rm RP}$ value of each star yields the conditional probability $P_{\rm GMM}(\mu_\alpha,\mu_\delta,\varpi | \alpha,\delta,G,G_{\rm BP}-G_{\rm RP})$. The contamination model can now be calculated as:
\begin{align}
\begin{split}
\mathcal{P}_{\rm cont}(\alpha,\delta,\mu_\alpha,\mu_\delta,G,G_{\rm BP}-G_{\rm RP},\varpi) = & \\
P_{\rm fine}(\alpha,\delta,G,G_{\rm BP}-G_{\rm RP}) \times & \\
P_{\rm GMM}(\mu_\alpha,\mu_\delta,\varpi | \alpha,\delta,G,G_{\rm BP}-G_{\rm RP}) \, \, . \\
\end{split}
\end{align}

In our initial development of the software we realised that it was undesirable to retain dense sources such as globular clusters in the input catalog. Otherwise, if their proper motions are poorly constrained, we find disk-like regions of fake high likelihood of size $\sim L$ (the stream search length parameter). On the other hand if the cluster's proper motions are well measured, the central cluster regions dominate the stream signal out to a distance $\sim L$. To avoid these undesirable effects we therefore removed all stars in the {\it Gaia} catalogs (DR2 and EDR3) within two tidal radii of the globular clusters listed in \citet{2010arXiv1012.3224H}, within 7 half-light radii of the Galactic satellites listed in \citet{2012AJ....144....4M}, and within a $3\deg$ radius of M31 and $1\deg$ radius of M33. We also excised sources near the (low-latitude) open clusters NGC~188, Berkeley~8, NGC~2204, NGC~2243, NGC~2266, Melotte~66, NGC~2420, NGC~2682 and NGC~6939 (using a suitable radius, typically $\sim 0.5\deg$).

The additional slight differences with respect to the analysis of Paper I are that here we consider distance solutions in the range $[1,30]\kpc$ (rather than just $[1,10]\kpc$), we consider SSP metallicity templates with ${\rm [Fe/H]= -2.2, -1.9, -1.5, -1.3, -1.1, -0.7}$ (rather than ${\rm [Fe/H]= -2.0, -1.6, -1.4}$), and we process the full sky (rather than just $|b|>20\deg$). Figure~\ref{fig:PMdist_all} shows the sky distribution of the 5960 stars selected for spectroscopic follow-up with a $7\sigma$ stream detection threshold (an $8\sigma$ threshold was used in Paper~I). As in paper~I, we adopt an extinction-corrected  limiting magnitude of $G_0=19.5$ for the search in DR2.

The sample shown in Figure~\ref{fig:PMdist_all} was selected to have $\sqrt{\mu_\ell^2+\mu_b^2} > 2 \masyr$, so as to avoid the contamination from low proper motion objects discussed at length in Paper~I. However, in order to avoid gaps that could be introduced by this proper motion selection, we inspected each stream and added in any of its members with $\sqrt{\mu_\ell^2+\mu_b^2} < 2 \masyr$. Sagittarius stream stars (identified as discussed in \citealt{2020ApJ...891L..19I}) were removed from this sample. Furthermore, at latitudes below $30\deg$, we selected features manually to optimize the spectroscopic follow-up campaign. However, all stars (above the chosen $7\sigma$ stream threshold) of a given stream or stream candidate were included. Above $|b| \simgt 30\deg$ the \texttt{STREAMFINDER} detections are unambiguous, but closer to the Galactic plane the algorithm finds high-significance stream-like correlated behavior in millions of {\it Gaia} stars. Since we do not have the resources to follow-up those potential detections, we took a pragmatic approach of selecting low-latitude structures by hand that were clearly well-separated from the bulk of the Galactic populations in the parameter space of observables (proper motion, parallax, color and magnitude). This approach allowed us to identify the low-latitude Hr\'{\i}d stream discussed below, as well as enabling us to follow previously-known structures in dense regions close to the Galactic plane. 

We expect the present stream sample to have uniform completeness for $|b|>30\deg$, but towards lower latitude the sample completeness diminishes, as we were progressively forced to select only the most significant structures. For $|b|<20\deg$, the sample is highly incomplete, and the selection is not remotely objective. However, the proof that the \texttt{STREAMFINDER} is providing useful detections is that all of the stream structures followed up spectroscopically to date have coherent line of sight kinematics.

Having processed the Gaia catalog with 6 different SSP model metallicity values, we select the solution that yields the highest likelihood, thus providing a distance and metallicity estimate for each star. Note that the distance is constrained by both the {\it Gaia} parallax and photometry (given the trial SSP model). The corresponding distance and metallicity solutions are shown in Figure~\ref{fig:dist_and_FeH}.

\begin{figure}
\begin{center}
\includegraphics[angle=0, width=\hsize]{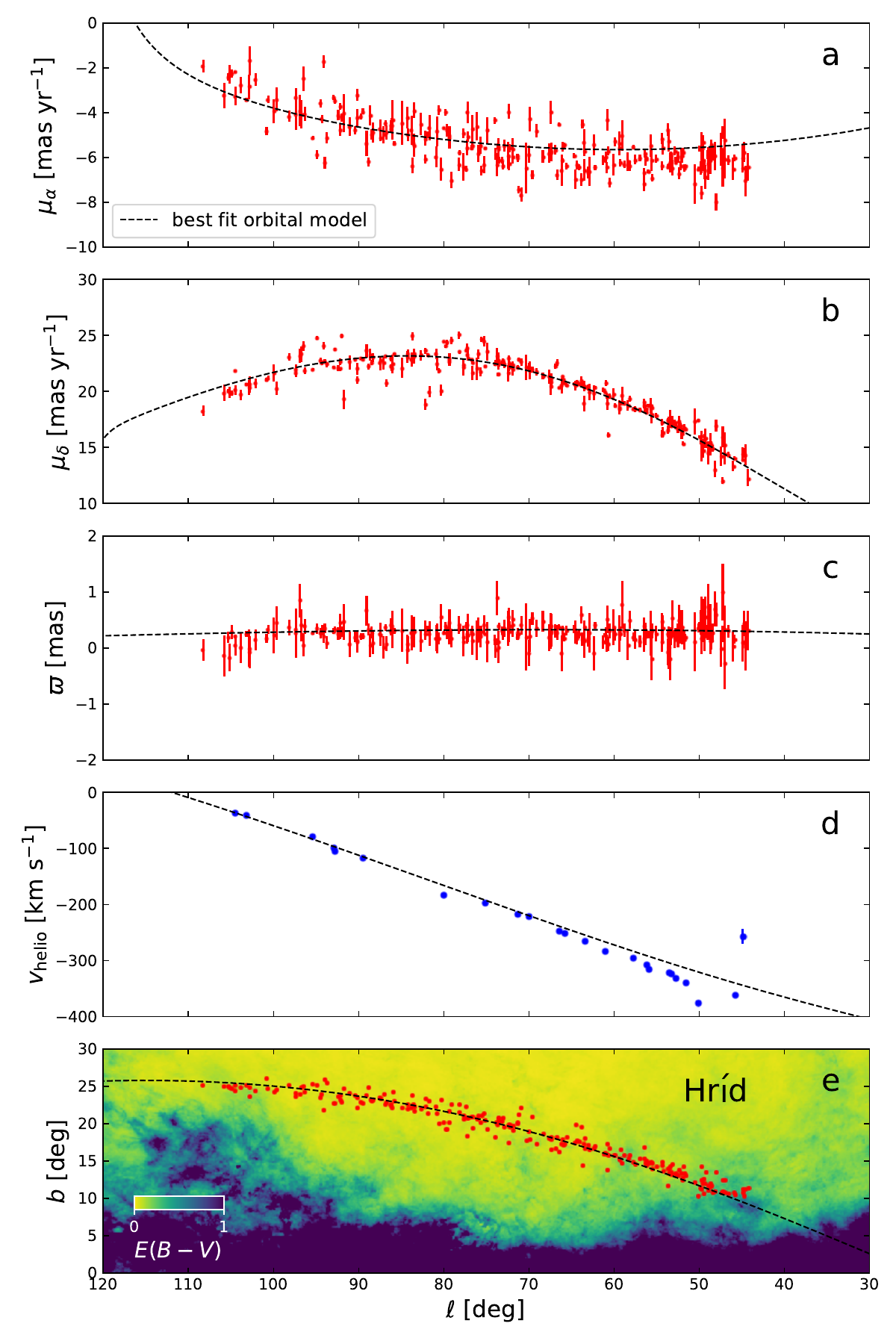}
\end{center}
\caption{Properties of the Hr\'{\i}d stream. From top to bottom, the panels show proper motion $\mu_\alpha$ and $\mu_\delta$, the parallax $\varpi$, the heliocentric line of sight velocity $v_h$ and Galactic latitude position of the stream stars, as a function of Galactic longitude $\ell$. The red points are the 156 members of the Gaia DR2 catalog with five-component astrometric solutions that are identified by the {\tt STREAMFINDER} software as stream stars using a $7\sigma$ threshold. The ($1\sigma$) uncertainties on these points are shown with error bars. The points in the fourth panel mark those stream members with measured radial velocity. As explained in the text, we fitted an orbit model to the astrometric and radial velocity data. The corresponding best-fit model is shown with a dashed line, and can be seen to give a reasonable representation of the parameter profiles. Additionally, the bottom panel also shows the $E(B-V)$ extinction map, extracted from \citet{Schlegel:1998fw}, from which the reader may verify that the detected stream does not follow any obvious structure in the interstellar extinction.}
\label{fig:Hrid}
\end{figure}

\begin{figure}
\begin{center}
\includegraphics[angle=0, width=\hsize]{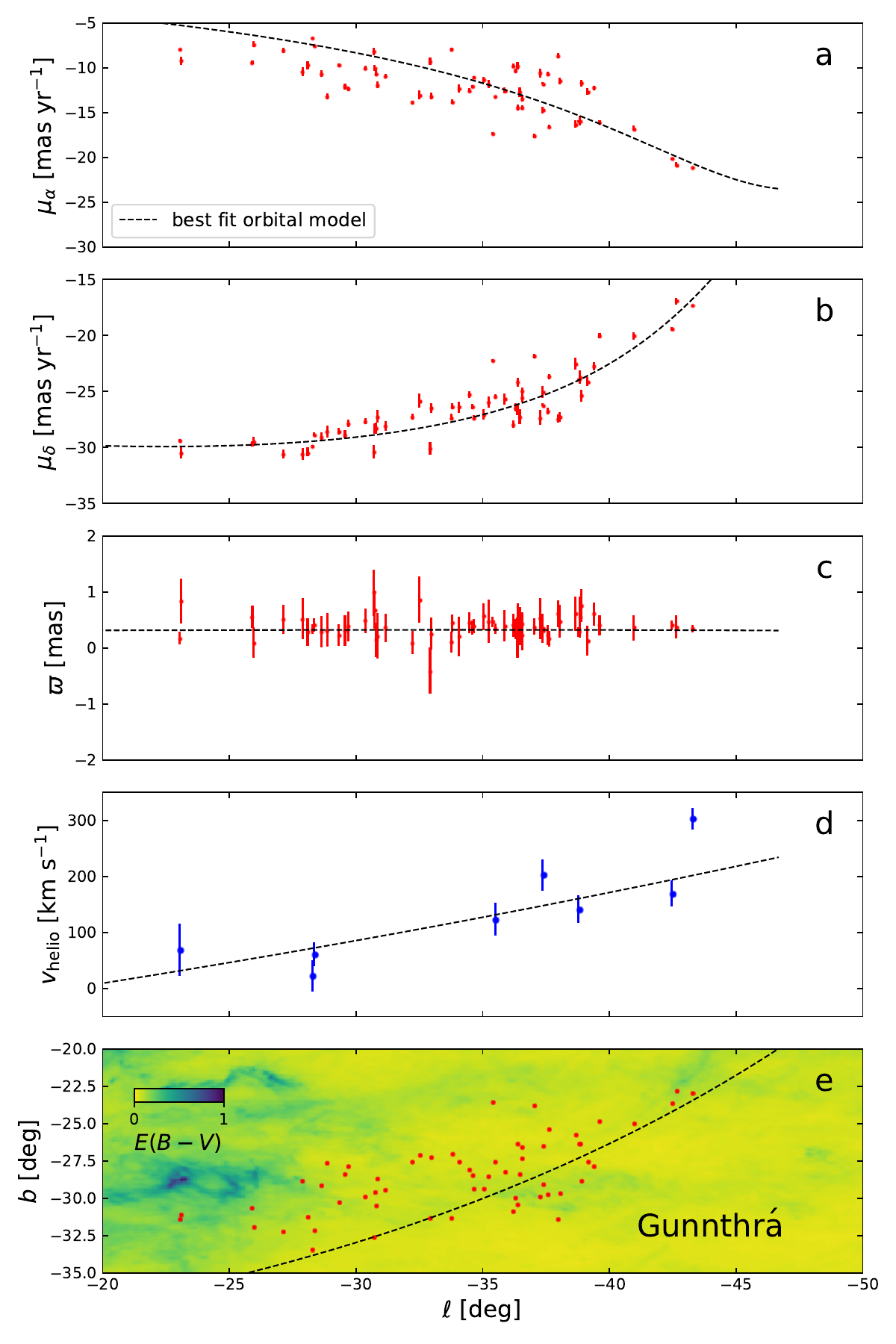}
\end{center}
\caption{As Figure~\ref{fig:Hrid}, but for the Gunnthr\'a stream. A total of 61 stars are identified as candidate members of this structure at $>7\sigma$ confidence.}
\label{fig:Gunnthra}
\end{figure}

\section{Spectroscopic Observations}
\label{sec:Observations}

To measure the radial velocities and metallicities of the DR2 sample presented above we secured 75 hours of service-mode observation with the CFHT/ESPaDOnS instrument  \citep{2003ASPC..307...41D} over five semesters between 2018 and 2020. We used ESPaDOnS in the ``object + sky'' non-polarimetric configuration, where it operates as a standard high-resolution spectrograph of resolution $R=68,000$ covering the wavelength range $370$ to $1,050$\AA. Our observing strategy was to aim for signal to noise of $S/N \sim 7$ per pixel for the majority of the fainter targets (up to $G_0=17.6$), while obtaining high signal to noise ($S/N>30$) for the brighter stars if it cost less than 20~min to do so. All the data were reduced with the Libre-ESpRIT pipeline \citep{1997MNRAS.291..658D} through to extracted and wavelength-calibrated one-dimensional spectra. A total of 163 stars were observed with ESPaDOnS up to the end of June 2020.

A further 6 nights of observation were obtained with the UVES spectrograph \citep{2000SPIE.4005..121D}, 3 in June 2019 and 3 in January 2020\footnote{Runs 0103.B-0568(B) and 0104.B-0406(B).}. Our aim with UVES was to probe fainter targets than can be efficiently observed with ESPaDOnS, and to reach regions of the southern sky that are inaccessible to the CFHT. UVES was used with the DIC2 dichroic beamsplitter in the ``437+760'' setting, covering 3730--4990\AA\ and 5650--9460\AA. For increased efficiency, we used $2\times2$ pixel binning when reading out the CCDs, and a $1\scnd$0 slit, which results in a resolution of $R\sim 40,000$. As with the ESPaDOnS observations, we calibrated the exposures to obtain sufficient signal to achieve $1$--$2\kms$ radial velocity accuracy for the fainter stars, while aiming for high S/N in the brightest targets, thus enabling chemical abundance studies. All the spectra were extracted and wavelength-calibrated with the ``esoreflex'' pipeline. Despite poor weather during the June 2019 run, we were able to measure the radial velocities of 174 stars with the UVES spectrograph over the two runs. (A final UVES run scheduled for April 2020 has been postponed to 2021 because of the Covid-19 pandemic).

As the UVES observations aimed to probe deeper than what we had managed to reach with ESPaDOnS, we considered it prudent to first check the velocities of the survey stars with a low resolution spectrograph before devoting precious VLT time to the follow-up. To this end we secured two 6-night runs with NTT/EFOSC2\footnote{Runs 0103.B-0568(A) and 0104.B-0406(A).} immediately before each UVES run. We used a narrow $0\scnd5$ slit together with grism \#19 to cover the spectral region 4400--5100\AA, yielding spectra with a resolution of $R\sim 3000$. The EFOSC2 observations were entirely reduced with the IRAF software\footnote{\url{http://ast.noao.edu/data/software}}. Most exposures were 20~min long, resulting in typical radial velocity uncertainties of $\sim 10\kms$. However, we quickly realised that the stream sample did not suffer from substantial contamination, and adapted our observation strategy accordingly. A total of 60 stars of the stream sample have an EFOSC2 velocity measurement but no UVES velocity measurement (either due to insufficient time to complete the follow-up or because the EFOSC2 measurement indicated that the star was unlikely to be a stream member). 

The radial velocities of all the stars in the sample observed with ESPaDOnS, UVES or EFOSC2 were measured using the \texttt{fxcor} algorithm in IRAF, by cross-correlation against the spectrum of the radial velocity standard star HD~182572. 

We complemented these velocities with measurements taken from public spectroscopic surveys. We cross-matched the stream sample against APOGEE-2 \citep{2017AJ....154...94M}, finding 16 in common; we find 2 stars in common with the GALAH DR3 survey \citep{2018MNRAS.478.4513B}; 25 in common with the Gaia Radial Velocity Spectrometer (RVS) catalog; 80 in common with the LAMOST DR5 survey \citep{2012RAA....12.1197C}; 4 in common with the Radial Velocity Experiment (RAVE DR5) \citep{2017AJ....153...75K}; and 150 in common with SDSS/Segue survey \citep{2009AJ....137.4377Y}. A further 7 stars are present in the kinematic survey of Palomar~5 by \citet{2009AJ....137.3378O}, and 22 stars in the survey of \citet{2017ApJ...842..120I}.

In those cases with multiple velocity measurements, we adopt the velocity value with the lowest uncertainty. In this way, a total of 685 stars from our sample of 5960 stream candidate stars (i.e. 11.5\%) end up having velocity measurements. Of these, 264 have velocity uncertainties $\sigma_v<1\kms$, 355 have $\sigma_v<2\kms$, 393 $\sigma_v<3\kms$, 491 have $\sigma_v<5\kms$, and 602 have $\sigma_v<10\kms$. The stars with measured velocities are displayed in color in Figure~\ref{fig:map_vels}a. Each marked stream shows up as grouping of stars of similar radial velocity, or as a feature with a position-dependent velocity gradient. In Figure~\ref{fig:map_vels}b we have assigned different colors to the various streams to allow the reader to distinguish the individual structures more easily.

The \texttt{STREAMFINDER} catalog of 5960 sources derived from {\it Gaia} DR2, but with updated {\it Gaia} EDR3 information is provided in Table~\ref{tab:data}. The radial velocity information is also listed. The catalog of new detections in EDR3 will be published when the full search is completed (the low-latitude disk and bulge region take millions of CPU hours of computation).

\begin{table*}
\caption{The first 10 rows of the {\tt STREAMFINDER} catalog of 5960 stars detected in {\it Gaia DR2},  with updated EDR3 astrometry.}
\label{tab:data}
\hskip -3cm
\begin{tabular}{lcccccccccccc}
\hline
\hline
EDR3 ID & $\alpha$ & $\delta$ & $\varpi$ & $\mu_\alpha$ & $\mu_\delta$ & $G_0$ & $(G_{BP}-G_{RP})_0$ & $d_{SF}$ & $v_h$ & $\delta v_h$ & s & S \\
 & $\deg$ & $\deg$ & $\mas$ & $\masyr$ & $\masyr$ & mag & mag & $\kpc$ & $\kms$ & $\kms$ &  &  \\
4976492500371673600 &     0.555095 &   -50.846349 &    -0.155 &     8.388 &    -4.010 &    19.452 &     0.668 &     9.333 &           &           &     &   1 \\
4973150156822139264 &     1.515135 &   -51.976175 &     0.174 &     6.889 &    -3.500 &    18.507 &     0.697 &    11.940 &           &           &     &   1 \\
4973592611470354688 &     1.773110 &   -50.834348 &     0.140 &     7.972 &    -3.240 &    19.180 &     0.582 &    10.044 &           &           &     &   1 \\
2429494258672369152 &     1.875748 &    -8.429741 &    -0.075 &     0.264 &     0.520 &    19.235 &     0.794 &     7.041 &           &           &     &   2 \\
4973559626121475968 &     2.854625 &   -50.733300 &     0.430 &     7.136 &    -3.877 &    19.429 &     0.574 &    10.274 &           &           &     &   1 \\
4973378962616486912 &     3.111449 &   -50.615773 &     0.121 &     7.975 &    -3.357 &    17.038 &     0.897 &    11.033 &    27.900 &     0.430 &  13 &   1 \\
2424367510829683456 &     3.216603 &   -12.313105 &    -0.002 &     0.174 &    -0.165 &    18.903 &     0.659 &     7.205 &           &           &     &   2 \\
2424399499746804608 &     3.824391 &   -12.055727 &    -0.147 &    -0.136 &     0.055 &    18.853 &     0.706 &     6.833 &           &           &     &   2 \\
2424146577712223616 &     3.963230 &   -12.643203 &     0.041 &    -0.078 &     0.012 &    17.946 &     0.622 &     6.922 &           &           &     &   2 \\
2417215325130569216 &     4.717180 &   -13.883024 &    -0.178 &    -0.763 &    -0.479 &    19.269 &     0.659 &     6.833 &           &           &     &   2 \\
\hline
\hline
\end{tabular}
\tablecomments{Column 1 provides the {\it Gaia} EDR3 identification of the star, 2--6 list the EDR3 equatorial coordinates $\alpha$ and $\delta$, parallax $\varpi$ and proper motions $\mu_\alpha (* \cos(\delta))$, $\mu_\delta$. The DR2 extinction corrected magnitude $G_0$, and color $(G_{BP}-G_{RP})_0$ used in the {\tt STREAMFINDER} are listed in columns 7 and 8, while column 9 provides the distance to the star $d_{SF}$ estimated by the algorithm. Columns 10 and 11 list the best measured heliocentric line of sight velocity, as derived from the corresponding source ``s'' in column 12. The source identifications ``s'' are: 1=APOGEE, 2=GALAH, 3=Gaia RVS, 4=LAMOST, 5=RAVE, 6=SDSS, 7=BOSS, 8=ESPaDOnS (this work), 9=AAOmega (from \citealt{2017ApJ...842..120I}), 10=FLAMES (from \citealt{2017ApJ...842..120I}), 11=UVES (from \citealt{2009AJ....137.3378O}), 12=EFOSC (this work), 13=UVES (this work). Finally, column 13 provides a unique stream identification label 1--32 (the different colors of the streams in the bottom panel of Figure~\ref{fig:map_vels} are assigned using these identification labels).}
\end{table*}

\begin{figure}
\begin{center}
\includegraphics[angle=0, width=\hsize]{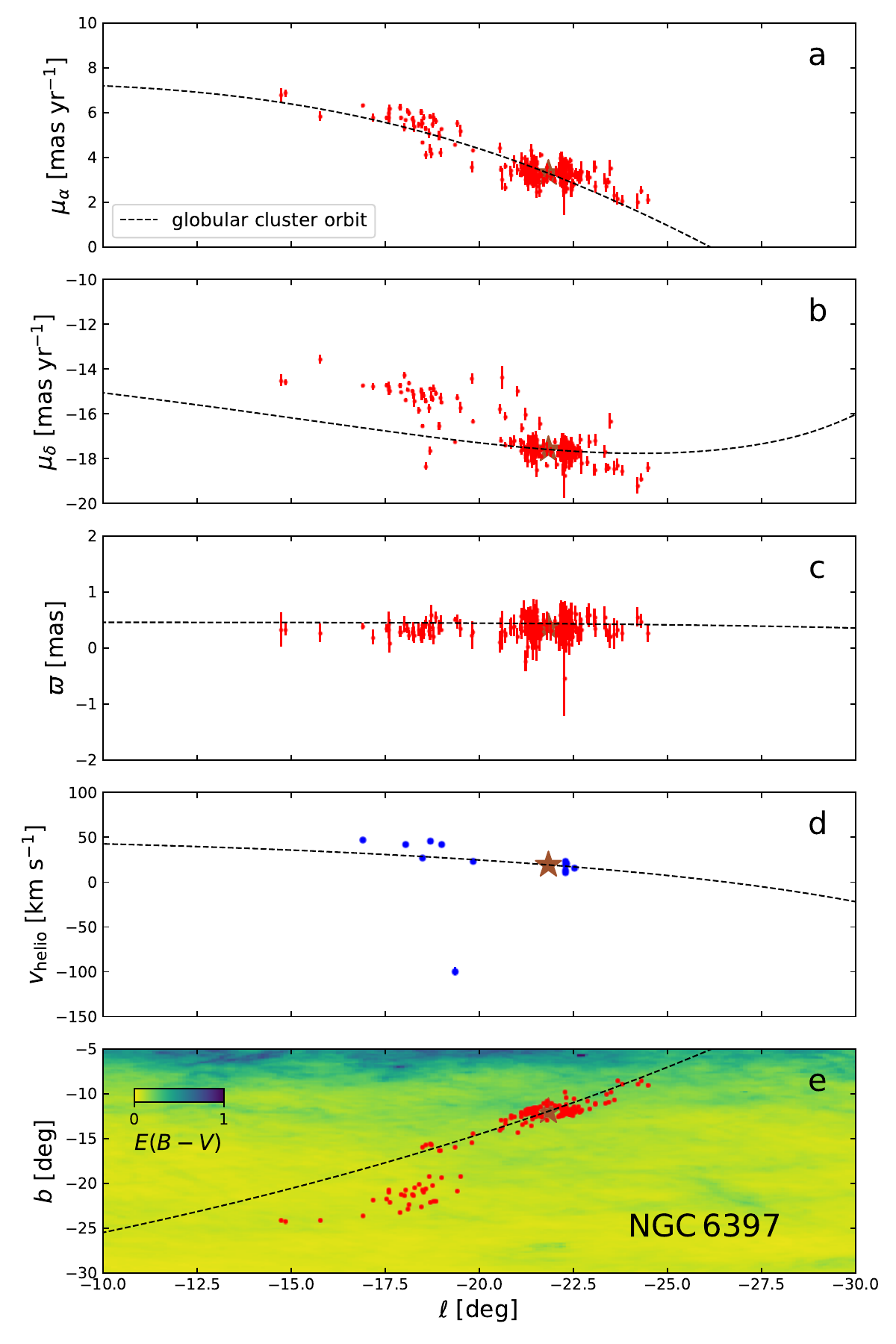}
\end{center}
\caption{As Figure~\ref{fig:Hrid}, but for the NGC~6397 stream. A total of 276 stars are identified as candidate members of this structure at $>7\sigma$ confidence. The globular cluster itself is represented with a star symbol, and the dotted line corresponds to the orbit of this object, as obtained by integrating using Gaia DR2 proper motions and position, the radial velocity and the heliocentric distance listed in \citet{2019MNRAS.484.2832V}. The Galactic potential model \#1 of \citet{Dehnen:1998tk} was used here.}
\label{fig:N6397}
\end{figure}

\begin{figure}
\begin{center}
\includegraphics[angle=0, width=\hsize]{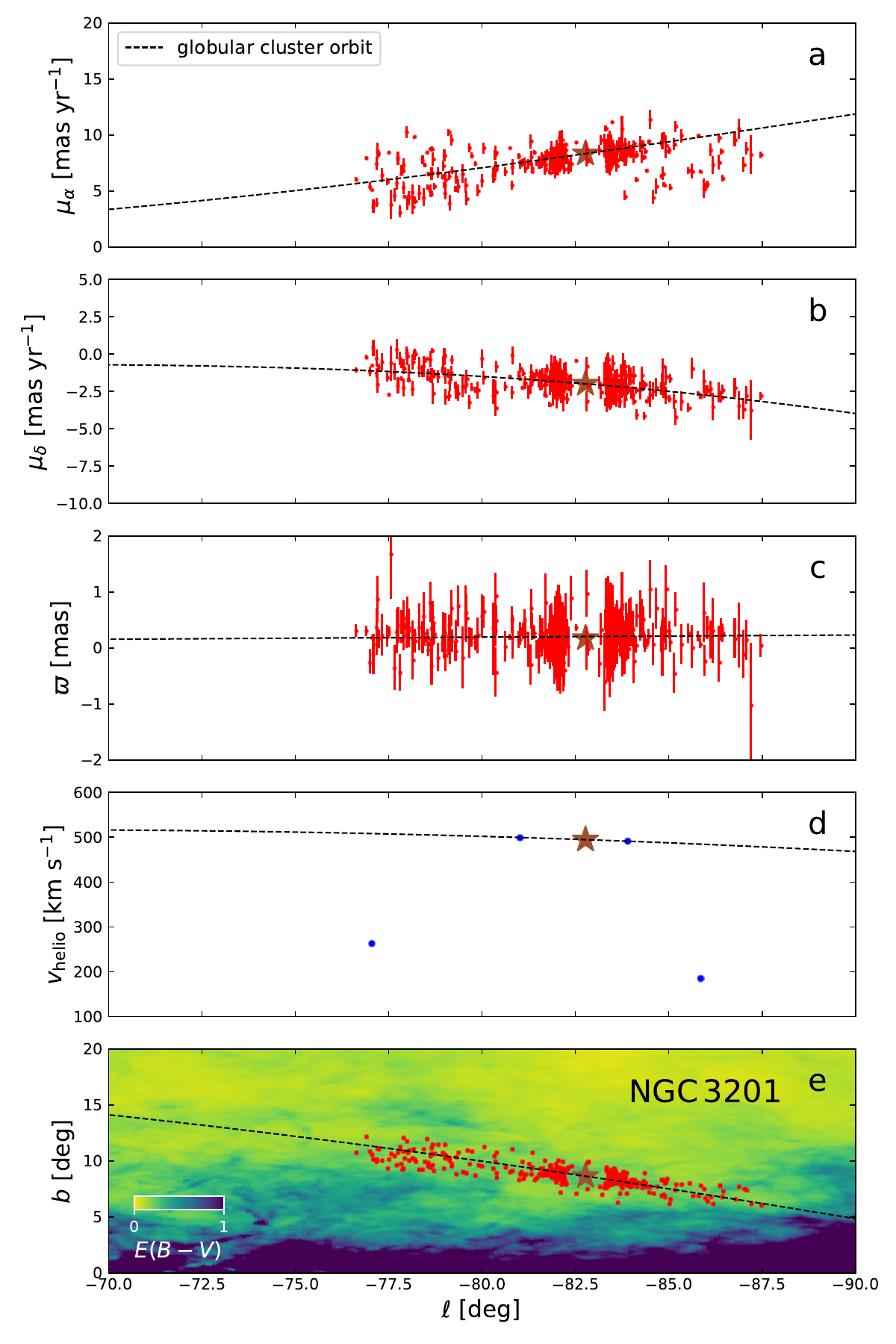}
\end{center}
\caption{As Figure~\ref{fig:N6397}, but for the NGC~3201 stream. A total of 388 stars are identified as candidate members of this structure at $>7\sigma$ confidence. }
\label{fig:N3201}
\end{figure}

\begin{figure}
\begin{center}
\includegraphics[angle=0, width=\hsize]{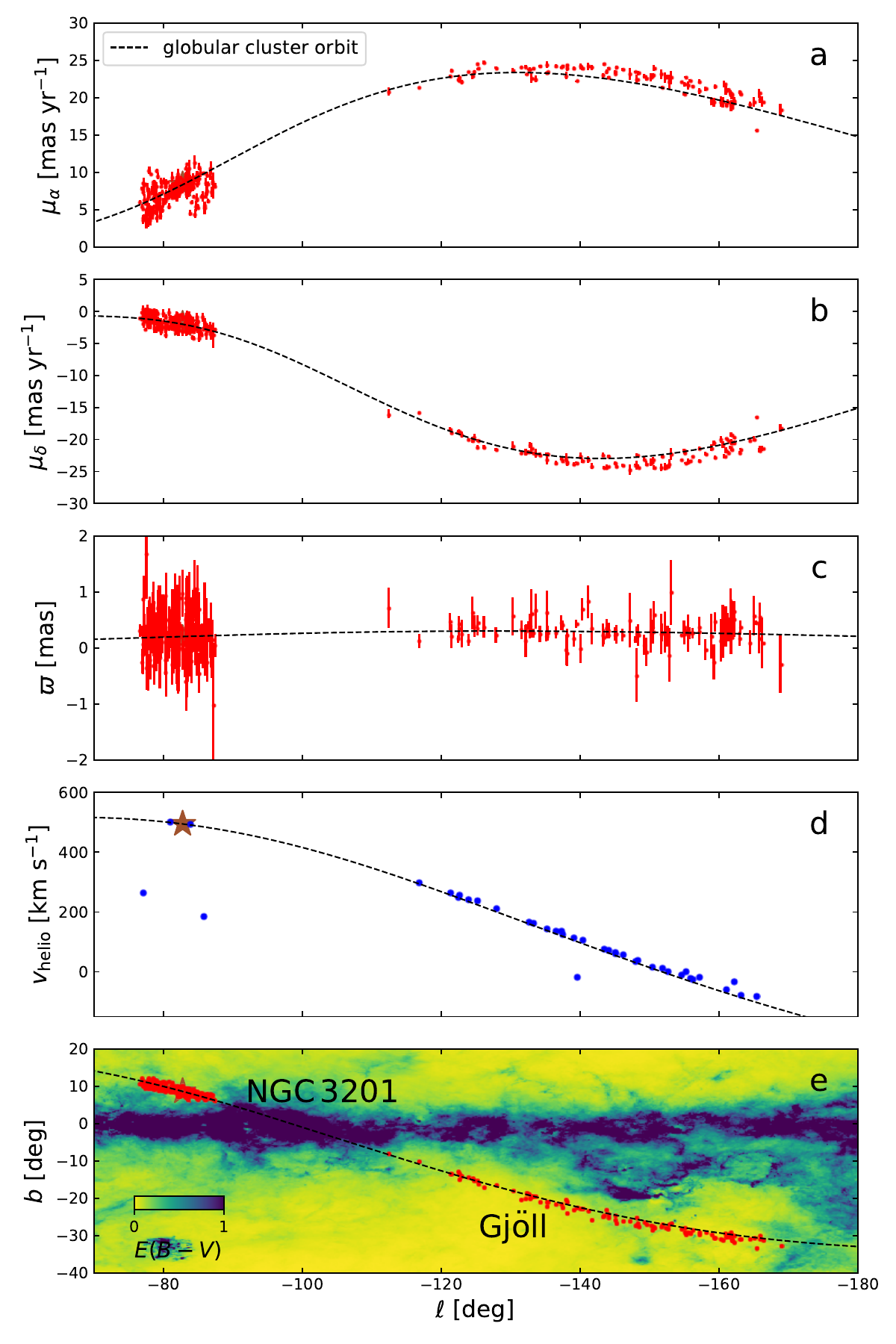}
\end{center}
\caption{As Figure~\ref{fig:N3201}, but showing a larger area of sky to encompass the Gj\"oll stream. The orbit shown is not a fit, but is rather simply the extrapolation from the measured phase-space values of NGC~3201. The excellent correspondence in position, distance, proper motion and radial velocity indicates that Gj\"oll is the trailing stream of NGC~3201.}
\label{fig:N3201_Gjoll}
\end{figure}

\section{{\it Gaia} DR2 Detections}
\label{sec:DR2_Detections}

We will first discuss the 32 streams or stream candidates detected in {\it Gaia} DR2, and presented in Figures~\ref{fig:PMdist_all}--\ref{fig:map_vels} and listed in Table~\ref{tab:data}. Some of these streams were previously known, or were announced over the course of our spectroscopic follow up campaign: GD-1 \citet{2006ApJ...643L..17G}, Orphan \citep{2006ApJ...642L.137B, 2006ApJ...645L..37G}, Palomar~5 \citep{2001ApJ...548L.165O}, ATLAS \citep{2014MNRAS.442L..85K}, Jhelum \citep{2018ApJ...862..114S}, Kwando\footnote{The distance solutions we find to Kwando ($\approx 8\kpc$) are substantially lower than the estimate of $\approx 20\kpc$ derived by \citet{2017ApJ...847..119G} from Pan-STARRS photometry. We consider the higher distance estimate to be more robust, since Pan-STARRS provides more accurate photometry than Gaia DR2 for the faint sources that make up the bulk of this structure.} \citep{2017ApJ...847..119G}, C-1 (which turns out to be the M~5 stream \citealt{2019ApJ...884..174G}), and M~92 \citep{2020ApJ...902...89T, 2020MNRAS.495.2222S}. The candidate streams C-1 to C-8 are structures that appeared real given extant astrometry and photometry, but which we could not confirm with complete confidence due to inexistent or insufficient follow-up observations. However, as we shall see below, EDR3 data has clarified the nature of some of them. The new stream structures presented in this section are the features named Hr\'{\i}d and Gunnthr\'a (following our previous nomenclature of adopting the names from Norse mythology of the streams of the gaping abyss that existed before the beginning of the world). We will also discuss the streams emanating from the globular clusters NGC~6397 and NGC~3201.

\subsection{Hr\'{\i}d}

Hr\'{\i}d is a low latitude stream that is very challenging to detect because it lies superposed on dense regions of the disk. We detect 156 members of this structure in {\it Gaia} DR2, and have 24 radial velocity measurements along it, including 8 UVES and 14 ESPaDOnS high precision measurements. Its properties are displayed in Figure~\ref{fig:Hrid}, where we show, as a function of Galactic longitude, the proper motion $\mu_\alpha$ (panel a), the proper motion $\mu_\delta$ (b), the parallax $\varpi$ (c), the heliocentric radial velocity $v_h$ (d) and its trend with Galactic latitude (e). We are able to follow the structure over $62\deg$ until it becomes lost in the high extinction at $b\sim 10\deg$. The distance estimates provided by the \texttt{STREAMFINDER} algorithm suggest that it lies at a heliocentric distance of $\sim 3.1\kpc$.

We fit an orbit to the astrometric and radial velocity data in the \citet{Dehnen:1998tk} Milky Way model `1' using a simple simplectic leapfrog integrator (the fitting procedure is identical to that outlined in \citealt{2018ApJ...865...85I}). The best fit orbit (dashed line) can be seen to follow closely the trends in each of the observed parameters. This orbit is extremely radial in the \citet{Dehnen:1998tk} potential: it has a pericenter of only $0.86\kpc$ and an apocenter of $34.0\kpc$. A dynamical study of this object could be interesting as it is hard to understand how it could have survived as a coherent structure until the present day.

The most likely metallicity found by the \texttt{STREAMFINDER} is ${\rm [Fe/H]=-1.1}$, which agrees well with the spectroscopic value of ${\rm [Fe/H]=-1.05\pm0.04}$ from one star in LAMOST and ${\rm \langle [Fe/H]\rangle=-1.13\pm0.04}$ from two stars in SEGUE. Thus it does not appear to be typical of the halo population. We will examine the constraints on the abundance distribution of this stream derived from the high-resolution spectroscopy in a subsequent contribution.

\subsection{Gunnthr\'a}

At first sight the $20\deg$-long Gunnthr\'a feature appears to be a continuation of the prominent retrograde Phlegethon stream (Figure~\ref{fig:EDR3PMdist_all}), as its stars have large proper motions that seem to continue the trend discerned in Phlegethon. However, the radial velocity of the member stars (Figure~\ref{fig:Gunnthra}d) are more than $200\kms$ lower than the trend observed in Phlegethon (also listed in Table~\ref{tab:data}).

All the spectroscopy of this stream is derived from the EFOSC2 spectrograph, and the spectra were not of sufficient quality to allow a metallicity measurement to be made. However, the \texttt{STREAMFINDER} finds the highest likelihood solutions with a metal-rich template of ${\rm [Fe/H]=-0.7}$. With this template the structure is estimated to be at a distance of $\sim 3\kpc$. The fitted orbit has a pericenter of $4.8\kpc$, an apocenter of $7.6\kpc$ and a maximum disk height of $4.0\kpc$, so it appears thick disk-like, but retrograde. This orbit is also quite diffefrent to that of Phlegethon (pericenter of $19.8\kpc$, and apocenter of $4.9\kpc$). Thus the alignment with Phlegethon on the sky and in proper motion is a coincidence.

Preliminary analysis of the low latitude extension of our survey suggests that the Gunnthr\'a feature may be the portion of a larger structure, possibly related to the Fimbulthul/NGC~5139 system. Additional analysis as well as spectroscopic follow up of the possible extension are ongoing to clarify the nature of this structure.

\subsection{NGC~6397}

One of the highlights of the search is the detection of the stellar stream of the nearest globular cluster NGC~6397 (Figure~\ref{fig:N6397}). The software finds stars from this cluster spread along an $18\deg$-long arc in that extremely low latitude field. Our spectroscopic follow-up of 8 stars with UVES confirmed the velocity membership, while of the 5 other stars found in public spectroscopic surveys only 1 is a velocity non-member. 

As far as we know, this is the first detection of an extended tidal stream associated to this nearby ($d_\odot=2.3\kpc$, \citealt{2010arXiv1012.3224H}), old (age$=12.6\pm 0.7\Gyr$, \citealt{2018ApJ...864..147C}) and metal-poor (${\rm [Fe/H]=-2.1}$, \citealt{2016A&A...588A.148H}) globular cluster.  The cluster is not included in the list by \citet{2020A&A...637L...2P}. \citet{2000A&A...359..907L} reported the detection of extra-tidal stars within $150\arcmin$ from the center of the cluster, with an hint of a tail reaching $\pm 100\arcmin$ from the center, approximately oriented along the E-W direction, in broad agreement with our findings. More recently \citet{2021A&A...645A.116K} used Gaia DR2 data to identify a handful of extra tidal stars within $5.0\deg$ of the cluster center. On the other hand, the stream detected here with Gaia DR2 data, albeit irregular, extends $18\deg$ on the sky.

\subsection{NGC~3201 and Gj\"oll}

NGC~3201 is also a low-latitude globular cluster that is suffering tidal stripping \citep{2019ApJ...887L..12B}. The \texttt{STREAMFINDER} is able to identify a $12\deg$-long feature in the immediate vicinity of the cluster, as we show in Figure~\ref{fig:N3201}. Unfortunately, we obtained only 4 velocity measurements in this region, 2 data from Gaia RVS and 2 UVES measurements. Of these 4 stars only 2 were confirmed as bona-fide members (panel d), but together with the astrometric data they constrain the orbit very well. Figure~\ref{fig:N3201_Gjoll} shows the result of extrapolating this orbit towards lower Galactic longitude $\ell$. We find an excellent correspondence with the Gj\"oll stream, previously detected in Paper~I. No additional fitting was performed. Clearly, Gj\"oll is the tidal stream of this globular cluster, spanning along a $98\deg$-long track on the sky. As pointed out already in Paper~I, this stream is retrograde.

\begin{figure*}
\begin{center}
\includegraphics[angle=0, viewport= 60 5 795 360, clip, width=\hsize]{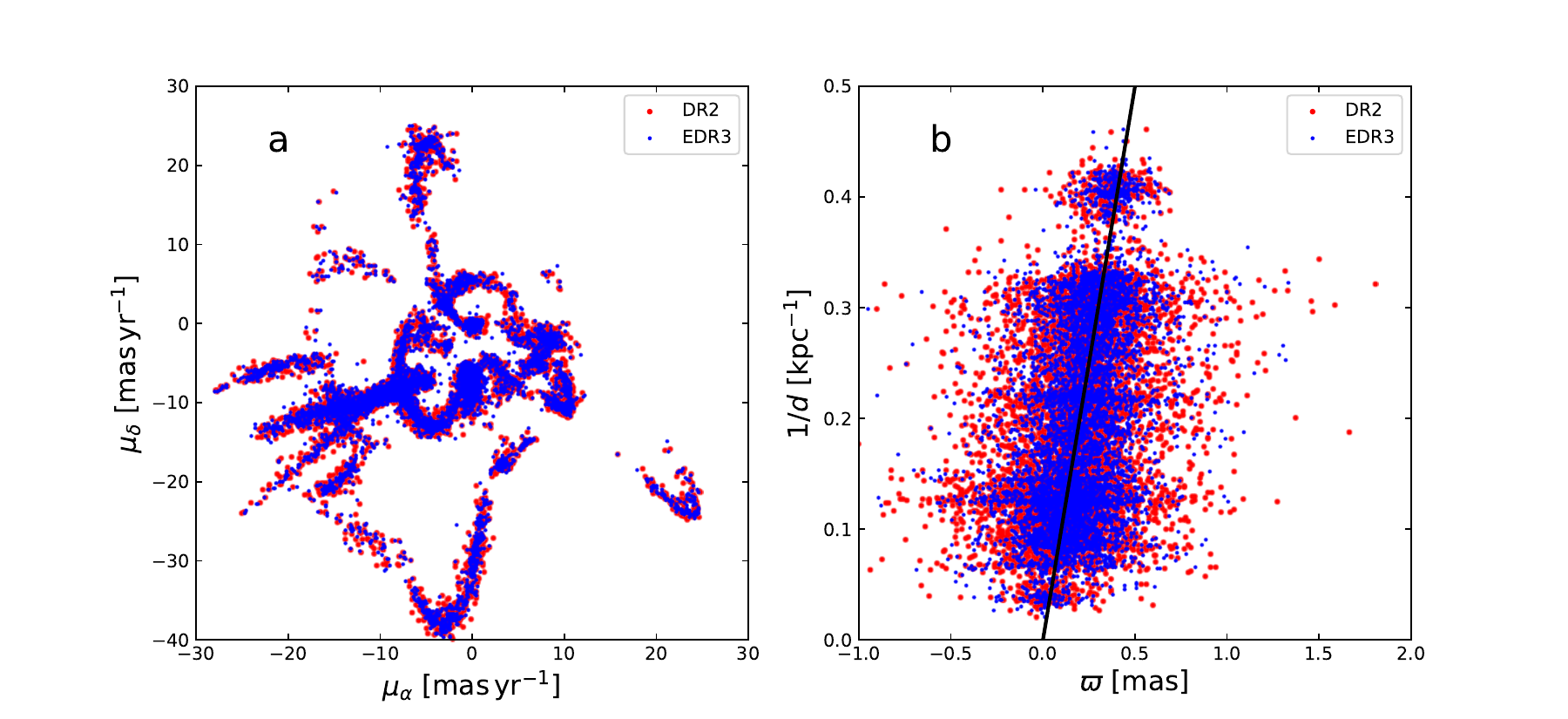}
\end{center}
\caption{Improvement in astrometric parameters between DR2 and EDR3. The proper motion distribution (a) of the sample of streams detected in DR2 is shown with the DR2 values (red) and EDR3 values (blue). There is a very clear reduction the proper motion scatter, which implied both that EDR3 has better proper motion measurements, and that the stream detection are real. The parallax values (b) also display a clearly improved scatter. (The solid straight line shows the expected one-to-one relation).}
\label{fig:DR2_DR3_comparison}
\end{figure*}

\begin{figure*}
\begin{center}
{\Large Gaia EDR3 detections, $[3,12]\kpc$}\par
\includegraphics[angle=0, viewport= 45 45 657 650, clip, width=16cm]{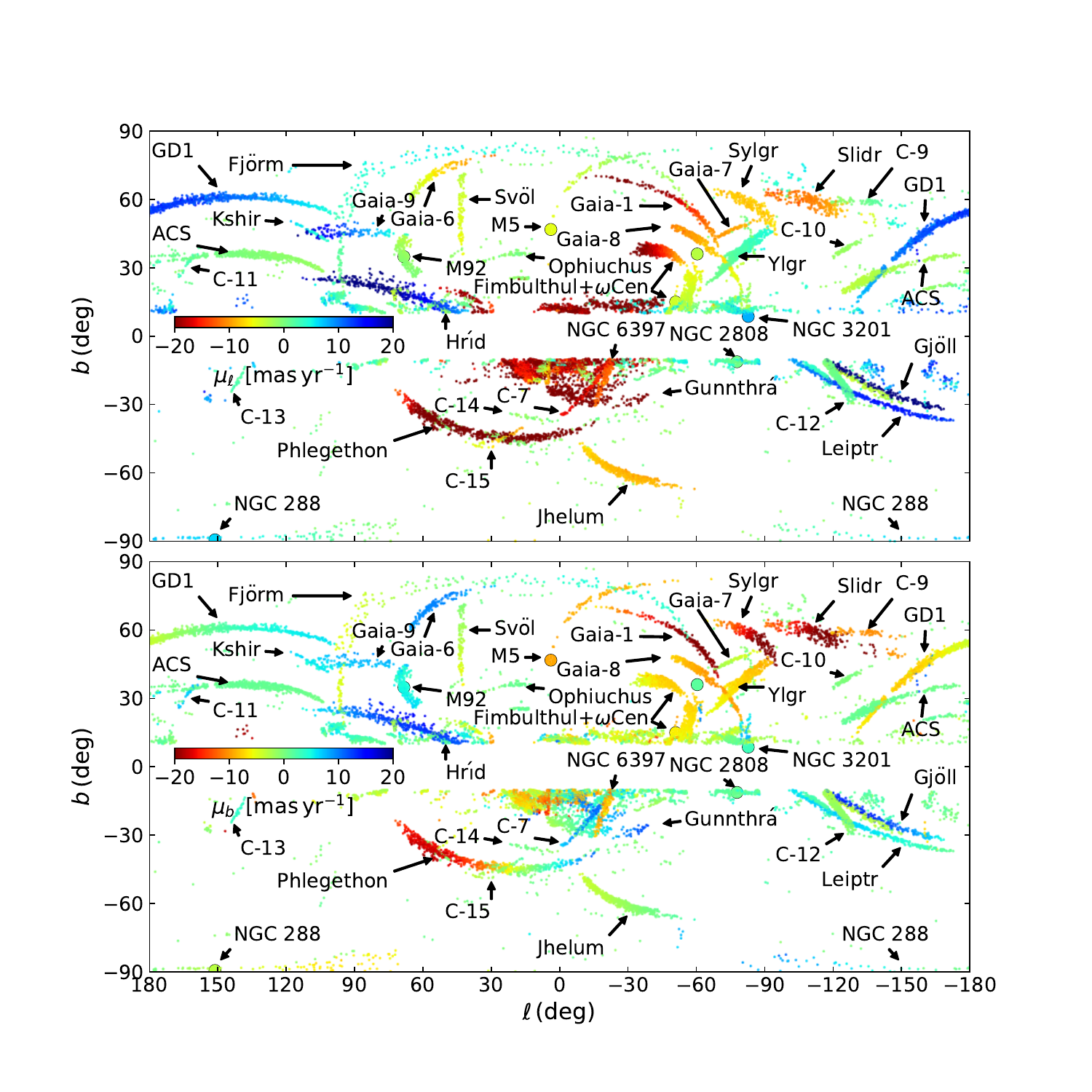}
\end{center}
\caption{New {\it Gaia} EDR3 map of the proper motion distribution in $\mu_\ell$  (top) and in $\mu_b$ (bottom) for the heliocentric distance range $[3,12]\kpc$. The sources are filtered so that only stars that have $>10\sigma$ likelihood of being stream members are shown. The globular clusters NGC~288, NGC~2808, NGC~3201, M~68, $\omega$Cen, M~5, M~92 and NGC~6397 are also shown, as their streams are visible on this map. The Anticenter Stream (marked ACS,  \citealt{2006ApJ...651L..29G, 2020arXiv201101241R}) appears in this map of thin streams, although note that the algorithm is finding a denser enhancement in that wide structure.}
\label{fig:EDR3PMdist_all}
\end{figure*}

\begin{figure}
\begin{center}
\includegraphics[angle=0, viewport= 40 10 495 385, clip, width=\hsize]{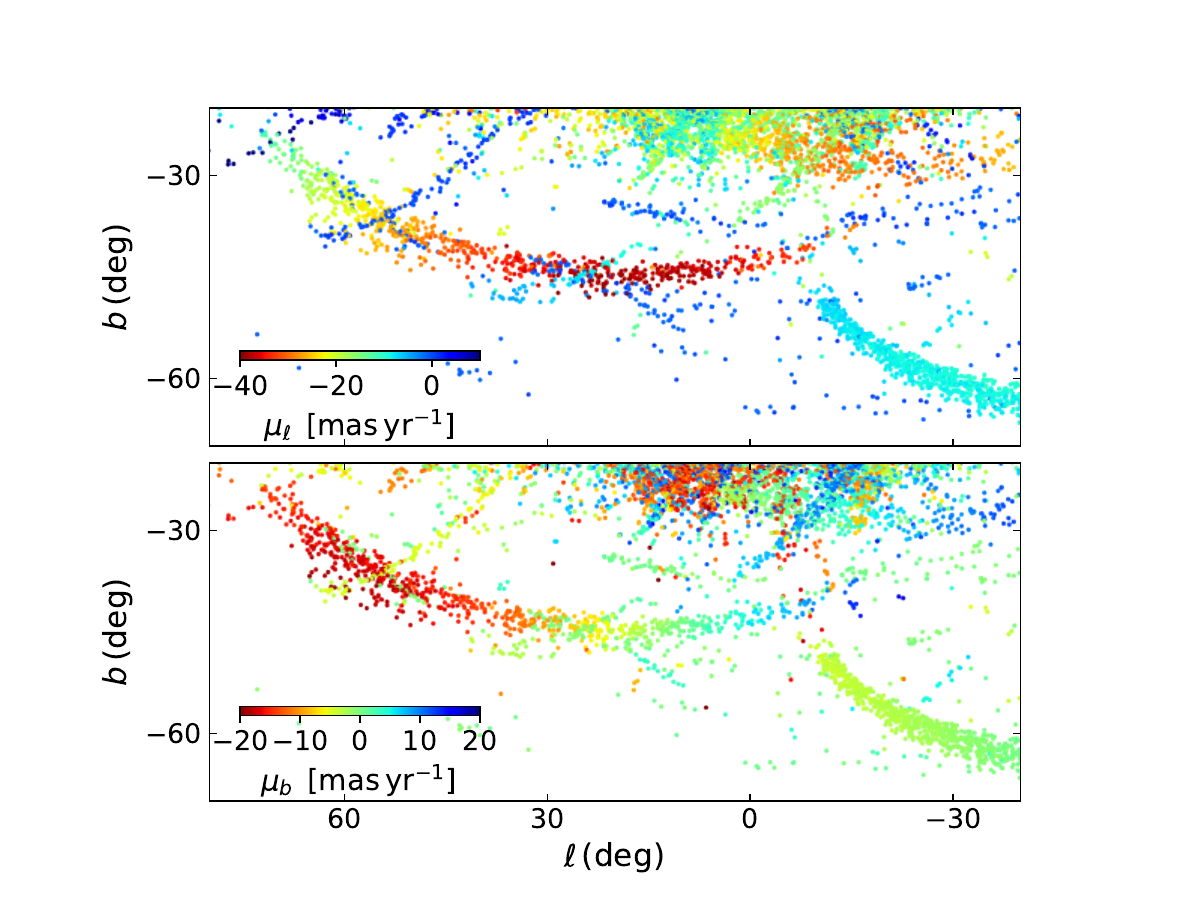}
\end{center}
\caption{As Figure~\ref{fig:EDR3PMdist_all}, but zoomed-in to the Phlegethon stream, and using a lower $8\sigma$ threshold. This region just south of the Galactic bulge is particularly rich in stream-like sub-structure, and it exemplifies the fact that many stream candidates become visible as we lower the detection threshold.}
\label{fig:EDR3PMdist_Zoom}
\end{figure}

\begin{figure*}
\begin{center}
{\Large Gaia EDR3 detections, $[10,30]\kpc$}\par
\includegraphics[angle=0, viewport= 45 45 657 650, clip, width=16cm]{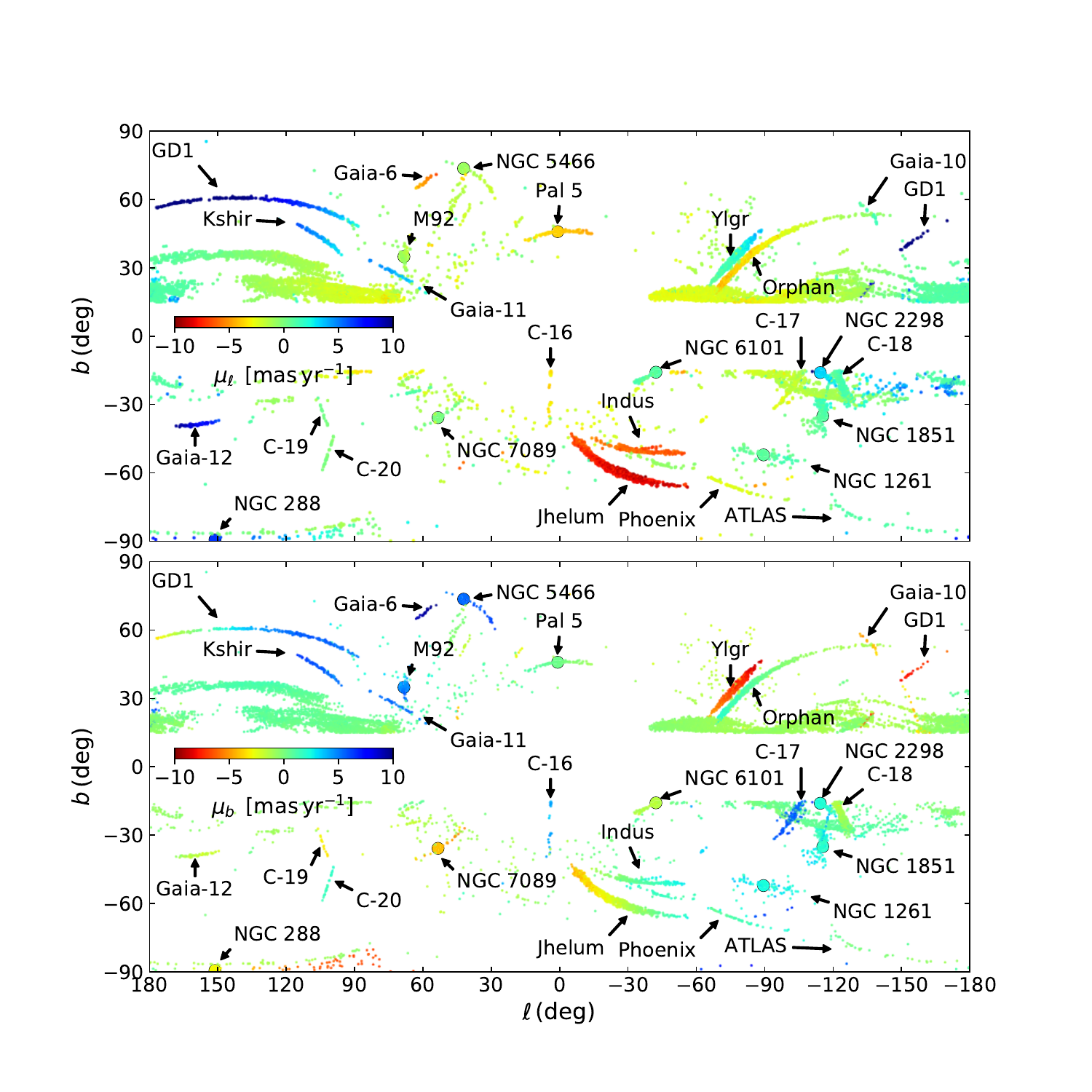}
\end{center}
\caption{New EDR3 map of the proper motion distribution in $\mu_\ell$  (top) and in $\mu_b$ (bottom) for the heliocentric distance range $[10,30]\kpc$. The sources are filtered to have $>10\sigma$ likelihood of being stream members, and to have metallicities ${\rm [Fe/H]<-1.2}$.}
\label{fig:EDR3PMdist_all_10to30}
\end{figure*}

\section{New Gaia EDR3 \texttt{STREAMFINDER} maps}
\label{sec:EDR3_maps}

The recent publication on December 3rd 2020 of the {\it Gaia} Early Data Release 3 catalog has allowed us to re-run the \texttt{STREAMFINDER} algorithm on this refined dataset. The procedure was essentially identical to that described in Section~\ref{sec:DR2_maps}, with a few exceptions. In EDR3 the parallax bias appears to be non-trivial to correct \citep{2020arXiv201201742L}, but the offset is tiny as far as the present work is concerned. So instead of the value of $0.029$~mas \citep{2018A&A...616A...2L} used in our DR2 search, we assumed a null parallax bias. As before, we used a template of age $12.5\Gyr$ and the metallicity values ${\rm [Fe/H]= -2.2, -1.9, -1.5, -1.3, -1.1, -0.7}$ were probed. To speed up the computation we undertook separate searches in two overlapping distance intervals: $[3,12]\kpc$ and $[10,30]\kpc$. The algorithm becomes computationally very expensive to run close to the Galactic plane, because of the large numbers of stars that need to be examined along each orbit. Because of this, we also decided to limit the present search to $|b|>10\deg$ for the $[3,12]\kpc$ interval and $|b|>15\deg$ for the $[10,20]\kpc$ interval. Given the increased astrometric and photometric precision of the EDR3 dataset, we also increased the limiting magnitude of the search from $G_0=19.5$~mag to $G_0=20$~mag. We expect later contributions in this series will provide updated maps that reach closer to the Galactic plane, and also perhaps to deeper limiting magnitudes.

However, we begin by showing the improvement that EDR3 brings to the problem of detecting stellar streams. In Figure~\ref{fig:DR2_DR3_comparison} we have cross-identified our DR2 stream catalog with EDR3 to obtain the new proper motion values. In panel (a), the new measurements (blue) are clearly less scattered than the old DR2 values (red). The increased accuracy of EDR3 should thus increase the contrast of the streams in proper motion space. Furthermore, the fact that the streams are tighter given better data provides further confirmation that the stream detections are real. We also show the improvement in the parallax of the sample in panel (b), plotted against the reciprocal of the distance estimated by the \texttt{STREAMFINDER} (from the DR2 data). Again there is a striking reduction in the parallax scatter, and it can also be seen that the algorithm produces distance estimates that are in broad agreement with the new parallax measurements.

The new sky map derived from EDR3 for the $[3,12]\kpc$ distance range is shown in Figure~\ref{fig:EDR3PMdist_all}. Similar to the DR2 case above, we again select stars that are not Sagittarius stream members, and reject sources with low proper motion so as to avoid contaminants. However, given the factor of $\sim 2$ improvement in the proper motions of the EDR3 catalog over DR2, this time we set the low proper motion cut to $\sqrt{\mu_\ell^2+\mu_b^2} > 1 \masyr$. Furthermore, to avoid having to display excessively crowded sky maps we increase the likelihood detection threshold to $10\sigma$. No other filtering has been applied (i.e. there has been no manual selection as in Figure~\ref{fig:PMdist_all}). The stream features previously seen in Figure~\ref{fig:PMdist_all} are also seen here, and several new structures are also apparent. In Figure~\ref{fig:EDR3PMdist_Zoom}, we show the effect of decreasing the detection threshold to $8\sigma$ in a particularly busy region that skims the southern bulge. A large number of kinematically coherent groups can be seen to be criss-crossing this field. The full sky view of the ($10\sigma$) streams in the $[10,30]\kpc$ distance window is shown in Figure~\ref{fig:EDR3PMdist_all_10to30}. Again, many new structures are revealed.

\section{{\it Gaia} EDR3 Detections}
\label{sec:EDR3_Detections}

The EDR3 maps (Figures~\ref{fig:EDR3PMdist_all}--\ref{fig:EDR3PMdist_all_10to30}) extend most of the streams seen in the DR2 maps (Figures~\ref{fig:PMdist_all}--\ref{fig:map_vels}). Several previously-known structures now come into view which were not present in Figure~\ref{fig:PMdist_all}: the M~5 stream \citep{2019ApJ...884..174G}, the Ophiuchus stream \citep{2014MNRAS.443L..84B}, the Anticenter Stream (ACS, \citealt{2003ApJ...594L.115R}), 
the Indus stream \citep{2018ApJ...862..114S} and the Phoenix stream \citep{2016ApJ...820...58B}. Seven new clear streams are also detected, which we name Gaia-6 to Gaia-12 (having ran out of names for Norse streams of the underworld). We also identify twelve candidate stream structures (C-9 to C-20), which appear plausible but require confirmation with follow-up observations.

\begin{figure*}
\begin{center}
\includegraphics[angle=0, width=16cm]{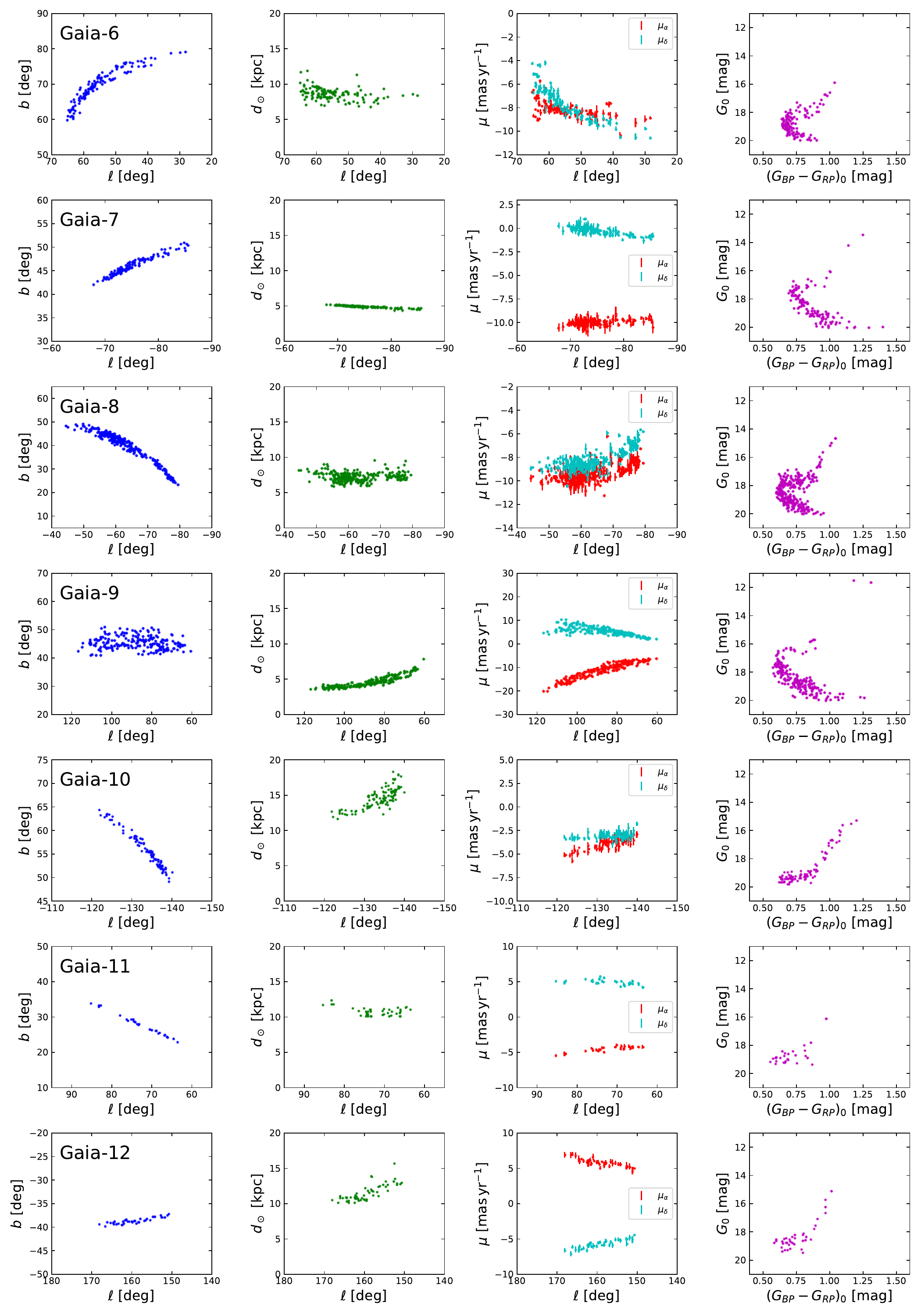}
\end{center}
\caption{Properties of the new streams Gaia-6 to Gaia-12 found in EDR3. The columns show, from left to right, the track of the stream on the sky, the distance profile as a function of Galactic longitude as estimated by the \texttt{STREAMFINDER}, the proper motion gradient in $\mu_\alpha$ (red) and $\mu_\delta$ (cyan), and the color-magnitude distribution of the group.}
\label{fig:Gaia_new}
\end{figure*}

Before proceeding to discuss the new detections, we note briefly that of the candidate structures marked in Figure~\ref{fig:PMdist_all}, we find that C-1 is a stream associated to M~5 \citep{2019ApJ...884..174G}, while C-2, C-3 and C-6 are clearly confirmed thanks to the excellent EDR3 data and are now named Gaia-9, Gaia-8 and Gaia-11, respectively. The position, estimated \texttt{STREAMFINDER} distance, proper motion and photometric properties of the new stream discoveries are summarized in Figure~\ref{fig:Gaia_new}.

\subsection{Gaia-6}

Gaia-6 is a $22\deg$-long feature that lies in the northern Galactic sky and appears strongest in the $[3,12]\kpc$ maps. The orbital solutions provided by the algorithm imply that it is situated at a distance of $d_{\odot}\sim 8 \kpc$. The very coherent proper motions in both $\mu_\alpha$ and $\mu_\delta$, the clearly defined color-magnitude distribution, and the fact that the object is detected at $15\sigma$ make this an unambiguous detection.

\subsection{Gaia-7}
Gaia-7 is a short stream that extends $14\deg$ in length. This stream is considerably closer to us, $d_{\odot}\sim 5 \kpc$, and it seems to possess negligible distance gradient. We find Gaia-7 to be highly coherent in proper motion, with a large magnitude of $\mu_\alpha$ that renders it easily distinguishable from the contamination. Given that this stream appears of highest contrast with the template model of ${\rm [Fe/H]=-0.7}$, it may well be that the progenitor of this stream was not a halo system. The highest likelihood regions of Gaia-7 are detected at $18\sigma$.

\subsection{Gaia-8}

Gaia-8 is another very interesting northern stream for which \texttt{STREAMFINDER} finds a large number of stars ($> 1000$ sources) over its $37\deg$ length. In the highest contrast regions it is detected at $14\sigma$ with the ${\rm [Fe/H]=-1.5}$ template. This stream lies at a distance of $d_{\odot} \sim 7\kpc$, and has a high measured proper motion value with an average of $\sim 12.5\masyr$.

\subsection{Gaia-9}
Gaia-9 runs for $40.5\deg$ roughly parallel to the Galactic plane. It appears as a thick stream in Figure~\ref{fig:EDR3PMdist_all}, partially due to its proximity, although it possesses a significant distance gradient ranging from $d_{\odot} \sim  3$--$7 \kpc$. This distance solution is based on a metal poor template of ${\rm [Fe/H]=-1.9}$. The tight proper motion profiles, as well as the color-magnitude distribution and $14\sigma$ detection, make this stream unambiguous.

\subsection{Gaia-10}
Gaia-10 is the farthest system that we have discovered in the present study, and it ranges from $d_{\odot}\sim 12-18\kpc$ according to the preferred template of metallicity ${\rm [Fe/H]=-1.9}$. This $17.6\deg$-long system is oriented almost perpendicular to the ``GD-1'' stream on the sky. We also detect several red giant branch stars for this stream (as can be seen from its color-magnitude distribution). The highest contrast regions of the structure are detected with $20\sigma$ confidence.

\subsection{Gaia-11}
Gaia-11 is a $19\deg$-long and extremely narrow structure, found at a distance of $d_{\odot}\sim  11\kpc$ with the ${\rm [Fe/H]=-1.9}$ template. Given its narrowness, we suspect the progenitor to be a globular cluster. The color-magnitude distribution of this stream is less well-defined than the other streams, but it is nevertheless detected at the $12\sigma$ confidence level. Future spectroscopic information should better inform us about the true nature of this stream.

\subsection{Gaia-12}
We find Gaia-12 as a $14\deg$-long stream in the southern Galactic sky. With the preferred ${\rm [Fe/H]=-1.9}$ metal-poor template, the system lies at $d_\odot\sim 11\kpc$, but with a clear distance gradient. Despite the low number statistics, the detection is unambiguous at $13\sigma$.

In future, we hope to undertake a careful investigation of the kinematics, stellar population and chemistry of these new ``Gaia~6--12'' streams, along with many other candidates that can be conspicuously seen in Figures~\ref{fig:EDR3PMdist_all}--\ref{fig:EDR3PMdist_all_10to30}.

\begin{figure*}
\begin{center}
\includegraphics[angle=0, viewport= 1 10 1000 1430, clip,width=16cm]{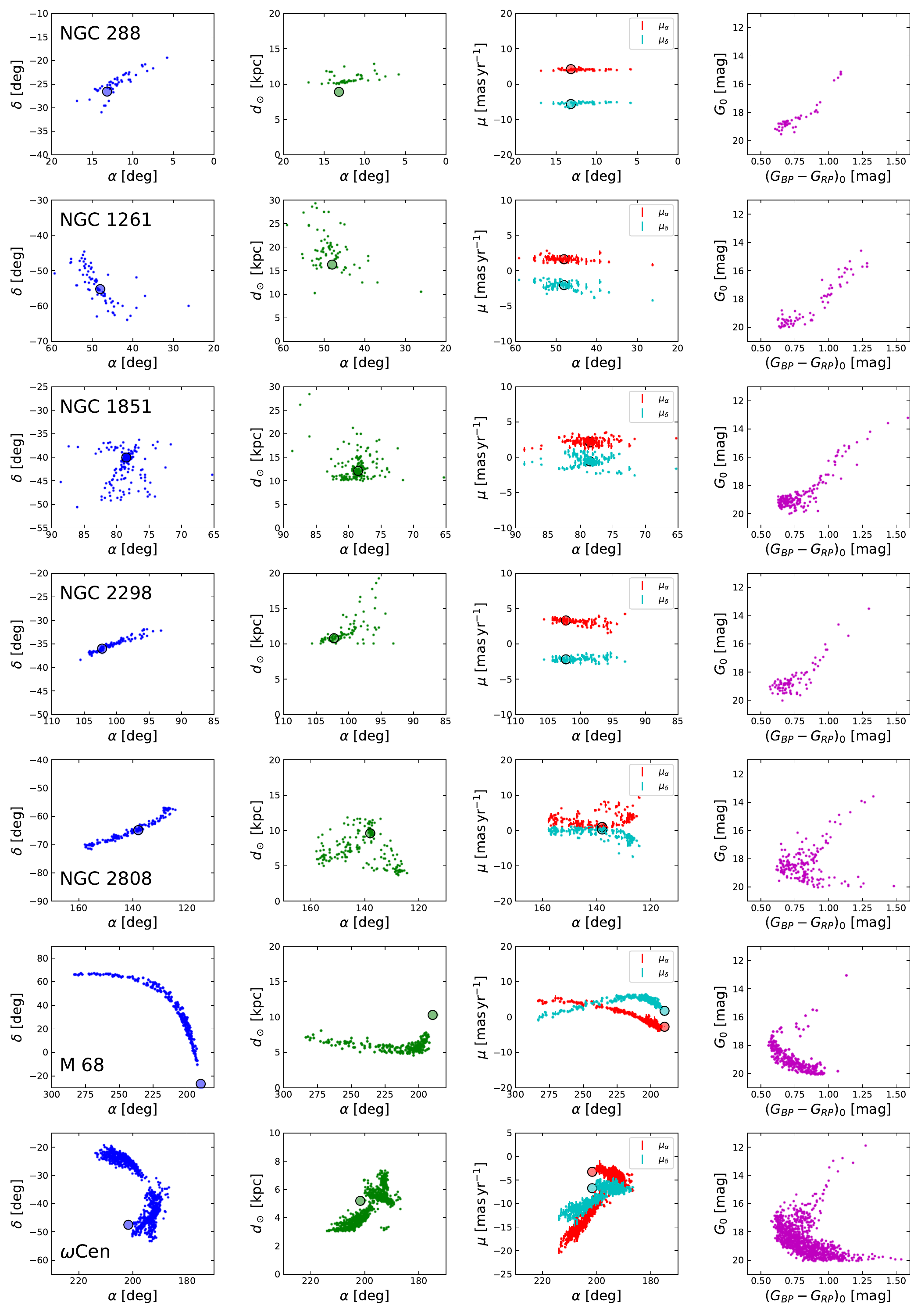}
\end{center}
\caption{Globular clusters with long tidal tails detected by the \texttt{STREAMFINDER}. The panels have the same layout as Figure~\ref{fig:Gaia_new}.}
\label{fig:Gaia_GCs1}
\end{figure*}

\begin{figure*}
\begin{center}
\includegraphics[angle=0, viewport= 1 180 1000 1430, clip,width=16cm]{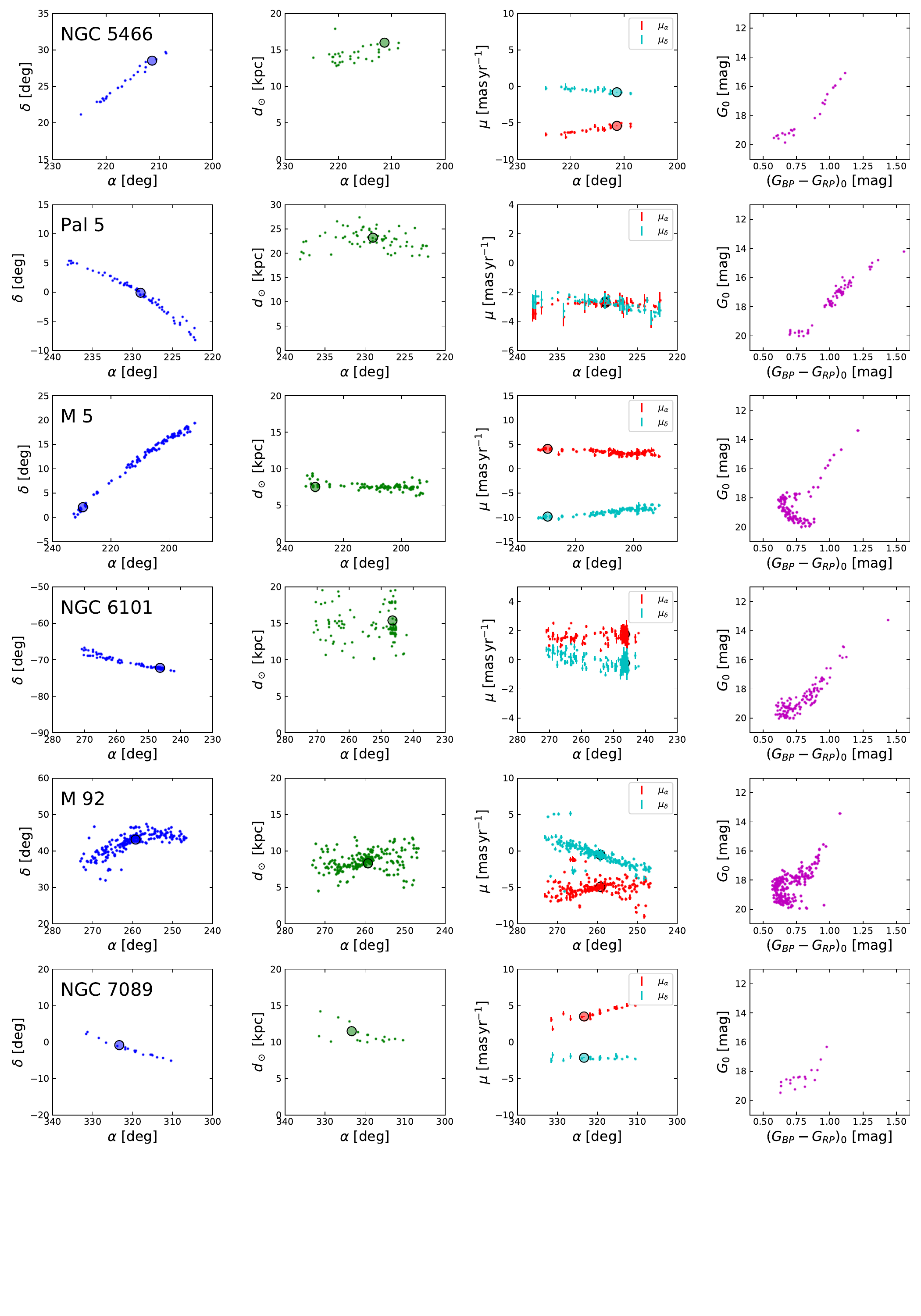}
\end{center}
\caption{Continuation of Figure~\ref{fig:Gaia_GCs1}.}
\label{fig:Gaia_GCs2}
\end{figure*}

\subsection{Globular clusters}

In the Gaia DR2 maps (Figure~\ref{fig:PMdist_all}), we found clear streams emanating from NGC~3201, M~68, $\omega$Cen, Palomar~5, M~92, and NGC~6397. The new processing of EDR3 maps also show long streams that are very likely associated to NGC~288, NGC~1261, NGC~1851, NGC~2298, NGC~2808, NGC~5466, M~5, NGC~6101, NGC~6397, and NGC~7089. In Figures~\ref{fig:Gaia_GCs1} and \ref{fig:Gaia_GCs2} we display all the stars in these structures derived from the {\it Gaia} EDR3 catalogs and that are identified as $>10\sigma$ by the algorithm (we do not include NGC~3201 or NGC~6397 which were shown previously in Figures~\ref{fig:N3201} and \ref{fig:N6397}, respectively). These clusters have good matches in position, distance and proper motion with extended  streams or fans that were found by the \texttt{STREAMFINDER}. In some cases, the stream is highly structured and further analysis and modelling is required to understand the structure and to disentangle the stream actually associated with the cluster from other unrelated structures that may cross that spot of sky just by chance. We defer the detailed analysis of these complex structures to a later contribution. We note that the present results on globular cluster streams are based on our \texttt{STREAMFINDER} analysis aimed at detecting stellar halo streams with a wide range of properties. We will do much better in future work by tailoring the stellar population templates to have the correct age and metallicity of each of the clusters. In particular, we expect this to improve the distance-metallicity degeneracy that affects the distance solutions found by the software.

For a comparison with the literature, in addition to our own search on each individual cluster, we took advantage of the recent work by \citep[][P20 hereafter]{2020A&A...637L...2P}, who scanned the literature on globular cluster tidal tails and extra-tidal stars, providing a final list of 53 clusters with reliable information on the presence or absence of extra-tidal stars. These authors classify the listed clusters into three groups: G1 for clusters with tidal tails, G2 for clusters with extra-tidal features, and G3 for clusters with no reported sign of extra-tidal stars in spite of dedicated studies. A systematic study on globular cluster tidal tails, based on Gaia DR2 data has been recently published by \citep[][S20 hereafter]{2020MNRAS.495.2222S}. This is a useful reference study to compare with, but only in the vicinity of the cluster, as the S20 analysis is limited to a circular area of radius $5\deg$ around the considered clusters.

NGC~288: The first detection of an extra-tidal extension in this cluster was reported by \citet{1995AJ....109.2553G}. More recently, \citet{2018ApJ...862..114S} presented photometry from the Dark Energy Survey that shows a possible detection of extra-tidal features over $\sim 8\deg$, oriented along the proper motion vector in the innermost 2--3$\deg$ and in the North-South direction beyond. In comparison, we trace the stream over $>10\deg$ degrees, approximately in the direction SE--NW, nicely aligned with the simulated orbit presented by \citet{2018ApJ...862..114S}. Additional detections within $2.0\deg$--$5\deg$ from the cluster center have been reported by \citet{2000A&A...359..907L}, \citet{2019MNRAS.484L.114K}, and S20. NGC~288 is classified as G1 by P20.

NGC~1261: The cluster is classified as G2 by P20, based on the detection of an extra-tidal envelope out to $r\simeq 0.5\deg$ from the cluster center by \citet{2018MNRAS.473.2881K}. \citet{2018ApJ...862..114S} report the possible detection of a $\sim 5\deg$ long tail, with similar orientation to the stream detected here. A photometric detection on small ($<1\deg$ deg) scale was reported by \citet{2000A&A...359..907L}.

NGC~1851: \citet{2018ApJ...862..114S} discussed the possible detection of a $\sim 8\deg$-long tail, approximately oriented as the inner part of the S-shaped tail detected here (as seen in equatorial coordinates). The detection of an extra-tidal envelope on the scale of $\la 2\deg$ was presented by \citet{2009AJ....138.1570O} and \citet{2018MNRAS.473.2881K}.
Possible spectroscopic detections were reported by \citet{2012MNRAS.426.1137S} and \citet{2014A&A...572A..30K}. The cluster is class G1, according to P20.

NGC~2298: Detections of an extra-tidal halo within $\simeq 1.0\degr$ from the center of this cluster, and, in some cases, of additional asymmetric components were reported by \citet{2000A&A...359..907L, 2011MNRAS.416..393B} and \citet{2018MNRAS.474..683C}. Here we present the detection of a coherent, filamentary structure spanning $\sim 12.0\degr$ on the sky, with an orientation that seems compatible with the tail found by S20, that extends to $\simeq \pm 5.0\degr$ from the cluster center. P20 class G2.

NGC~2808: We detect a very significant and narrow stream spanning about $\pm 10.0\degr$ around the cluster, with a substantial distance gradient. The structure is nearly parallel and almost adjacent to the low galactic latitude limit of our survey, still the match in position, distance and proper motion with the cluster suggests that the association is real. Extra-tidal stars were previously found in the surroundings of the cluster (within $0.5\degr$) by \citet{2018MNRAS.474..683C} and by \citet{2021A&A...645A.116K}, while no tail was found by S20.

NGC~3201: Very recently \citet{2020arXiv201014381P} detected the stream around the cluster and the associated Gj\"oll stream, in full agreement with our results. However, their detection of Gjoll is more extended than ours, reaching  $(\alpha,\delta)=(40\deg,+20\deg)$ instead of $(\alpha,\delta)=(68\deg,0\deg)$. These authors suggest that the gap between the two branches is due to an high extinction region intervening along the line of sight. A good chemical match between the cluster and Gj\"oll has been found by \citet{2020ApJ...901...23H}, based on the abundance analysis of four kinematic members of the stream.
See also \citet{2019ApJ...887L..12B}, S20 and \citet{2014A&A...572A..30K} for previous additional detections on small scales or based on small samples.

M68 (NGC~4590): \citet{2019MNRAS.488.1535P} associated this cluster with the Fj\"orm stream, providing a detection, based on Gaia DR2, fully consistent with that presented here. The cluster is P20 class G1, and the stream was detected also by S20.

$\omega$Cen: The most massive global cluster in the Milky Way has long been suspected to be the remnant of the central nucleus of an accreted galaxy \citep{2000LIACo..35..619M} because of its significant metallicity spread (e.g. \citealt{2010ApJ...722.1373J}) and because its kinematics vary as a function of abundance \citep{2018ApJ...853...86B}, which strongly suggest a protracted formation history. A well-populated $28\deg$ long tidal stream emanating from $\omega$Cen was detected in Gaia DR2 data \citep{2019NatAs...3..667I}, although at $b<25\deg$ the \texttt{STREAMFINDER} was unable to find the continuation of the stream, so a matched-filter technique was used instead with a further kinematic selection informed by an N-body model. The present Gaia EDR3 results are consistent with that earlier analysis associating the ``Fimbuthul'' stream with this cluster, but now with EDR3 we are able to directly detect member stars with the \texttt{STREAMFINDER} software to $b=10\deg$. In future work we will continue to search for the stream closer to the Galactic plane. P20 class G1.

NGC~5466: We find the same structure that was originally discovered by \citet{2006ApJ...637L..29B} and \citet{2006ApJ...639L..17G}. Here we trace the stream from $(\alpha,\delta)\simeq (225\deg,21\deg)$ to
$(\alpha,\delta)\simeq (212\deg,30\deg)$, just beyond the position of the cluster, while the detection by \citet{2006ApJ...639L..17G} extends to $\alpha\simeq 190\deg$ and possibly beyond (see \citealt{2012MNRAS.424L..16L}). P20 class G1.

Palomar~5: Ever since \citet{2001ApJ...548L.165O} first revealed the  massive tidal tails emanating from this object, Palomar~5 has been considered the ``poster child'' \citep{2015ApJ...803...80K} for the disruptive effect of tidal stripping on globular clusters. Using Gaia DR2 data \citet{2020MNRAS.493.4978S} detected this system over $\approx 27\deg$, extending the leading arm by $\approx 7\deg$ compared to the deep CFHT photometric mapping by \citet{2016ApJ...819....1I}. The present detection in Gaia EDR3 finds a $21\deg$ long structure with 10$\sigma$ stream members. The cluster is not listed by P20.

M5 (NGC~5904): \citet{2019ApJ...884..174G}, using Gaia DR2 data, detected an extended tail in excellent agreement with our finding. P20 class G1.

N6101: A highly significant narrow stream extending for about $10\degr$ in the sky. We are not aware of any previous detection. Not included in the P20 list.

NGC~6397: With Gaia EDR3 data, we find a well-populated stream over $25\deg$ in length, substantially extending earlier detections of extra-tidal stars by \citet{2000A&A...359..907L} and \citet{2021A&A...645A.116K}, within $100\arcmin$ and $5\deg$ of the cluster center, respectively. The cluster is not included in the list compiled by P20. Recent Jeans modelling of NGC~6397 suggests that it may contain a population of black holes at its core \citep{2021A&A...646A..63V}, while its survival to the present day without forming a conspicuous stream appears to be difficult to explain unless it has been protected by a dark matter sub-halo \citep{2021arXiv210403635B}. Our new observational constraints may help inform this debate.

M92 (NGC~6341): Despite the G3 class assigned to this cluster by P20, S20 detected the tail within 5 degrees from the center and \citet{2020ApJ...902...89T} traced the cluster stream with deep CFHT and Pan-STARRS photometry over 17 degrees. We are clearly detecting the structure here, approximately with the same extension as \citet{2020ApJ...902...89T}.

NGC~7089: This cluster presents an extended ($22\degr$) and narrow stream. It was not detected by S20, and it is not included in the P20 list. An extra-tidal component on small scales ($< 2.0\degr$) was reported by \citet{1995AJ....109.2553G}.

\section{Discussion and Conclusions}
\label{sec:Conclusions}

We have embarked on a large program to detect and characterize the stellar streams of the Milky Way so as to finally employ them as a means to constrain the properties of dark matter and compare the merits of different theories of gravity. The first step of this endeavor was the development of an optimized search tool, the \texttt{STREAMFINDER} \citep{2018MNRAS.477.4063M} to hunt through astrometric datasets. Initial results of the application of that algorithm to Gaia DR2 were reported in \citet{2018MNRAS.481.3442M} and \citet{2019ApJ...872..152I}.

In this contribution we first presented the stream sample (Figure~\ref{fig:PMdist_all}) that we constructed as the foundation for a high-resolution spectroscopic follow-up campaign conducted from 2018--2020 with CFHT/ESPaDOnS in the northern hemisphere and ESO/UVES in the south. The source catalog of 5960 stars, along with the radial velocity measurements for 685 of this set, are provided in Table~\ref{tab:data}. We find that all of the detected streams have coherent kinematics, as expected.

We have taken the opportunity of the very recent publication of the EDR3 catalog of the {\it Gaia} mission to update further our view of the ancient accretions onto the Milky Way. EDR3 provides a significant improvement in astrometric quality over DR2, which causes streams to be tighter and hence higher-contrast features in the parameter space of observables (Figure~\ref{fig:DR2_DR3_comparison}). 

We find nine new streams without an obvious progenitor (2 in our DR2 search, and 7 additional ones in the EDR3 survey), ranging in distance from $\sim 3\kpc$ to $\sim 20\kpc$. Our choice to filter sources with $\sqrt{\mu_\ell^2+\mu_b^2}<1\masyr$ undoubtedly biasses us against finding very distant structures. The filter was applied to avoid contamination from the numerous sources that appear close to being at rest. We have already upgraded the algorithm \citep{2020ApJ...891..161I} to be able to use ancillary multi-band photometric data (such as from Pan-STARRS), which may allow us to avoid the proper motion filter and thus search for more distant streams with our technique. This will be investigated in future work.

Since the pioneering work of \citet{1995AJ....109.2553G} and \citet{2000A&A...359..907L}, an ever-increasing list of globular clusters have been found to possess tidal tails (e.g., \citealt{2001ApJ...548L.165O, 2019NatAs...3..667I, 2019MNRAS.488.1535P, 2019ApJ...887L..12B, 2020ApJ...902...89T, 2020MNRAS.495.2222S}). The present work reveals that at least 15 clusters have exceedingly long tidal tails with numerous member stars measured with exquisite astrometric precision in the {\it Gaia} DR2 and EDR3 catalogs. These structures are clearly ideal candidates for probing the Galactic potential, especially since it will be possible to anchor their absolute distances to excellent accuracy, and the relative distances can also be measured very well because the cluster provides a perfect stellar populations template for the stream. They are (in order of increasing right ascension of the remnant cluster): NGC 288, NGC~1261, NGC~1851, NGC~2298, NGC~2808, NGC~3201, M~68, $\omega$Cen (the most massive globular cluster), NGC~5466, Palomar~5, M~5, NGC~6101, M~92, NGC~6397 (the closest globular cluster), and NGC~7089. Studying how these clusters managed to survive until the present day in the complex environment of the Milky Way may place interesting constraints on the dynamical evolution of the clusters and the mass distribution of their host.

{\it Gaia} has opened up a spectacular vista onto the Milky Way's halo, revealing a lacework of criss-crossing ancient and ongoing accretions that testify to the violent formation history of our Galactic home. As we turn from the halo towards the inner Galaxy, the weave becomes yet more complex, resembling a ball of wool (Figure~\ref{fig:EDR3PMdist_Zoom}). It will clearly take some effort to disentangle these structures, but now, finally with {\it Gaia} such an endeavor is becoming possible. In future contributions in this series, we will attempt to decipher this rich trove, presenting the detailed chemical abundances of the streams, developing N-body models of each structure, and performing a conjoint analysis of their constraints on the large- and small-scale Galactic acceleration field.

\acknowledgments

RI, NM, DA, BF, GM and AS acknowledge funding from the Agence Nationale de la Recherche (ANR project ANR-18-CE31-0006, ANR-18-CE31-0017 and ANR-19-CE31-0017), from CNRS/INSU through the Programme National Galaxies et Cosmologie, and from the European Research Council (ERC) under the European Unions Horizon 2020 research and innovation programme (grant agreement No. 834148). FR acknowledges support from the Knut and Alice Wallenberg Foundation. MB acknowledges the financial support to this research by INAF, through the Mainstream Grant 1.05.01.86.22 assigned to the project “Chemo-dynamics of globular clusters: the Gaia revolution” (P.I. E. Pancino). GT acknowledge support from the Agencia Estatal de Investigaci\'on (AEI) of the Ministerio de Ciencia e Innovaci\'on (MCINN) under grant with reference (FJC2018-037323-I).

We gratefully acknowledge the High Performance Computing centre of the Universit\'e de Strasbourg for a very generous time allocation and for their support over the development of this project.

This work has made use of data from the European Space Agency (ESA) mission {\it Gaia} (\url{https://www.cosmos.esa.int/gaia}), processed by the {\it Gaia} Data Processing and Analysis Consortium (DPAC, \url{https://www.cosmos.esa.int/web/gaia/dpac/consortium}). Funding for the DPAC has been provided by national institutions, in particular the institutions participating in the {\it Gaia} Multilateral Agreement. 

Based on observations obtained at the Canada-France-Hawaii Telescope (CFHT) which is operated by the National Research Council of Canada, the Institut National des Sciences de l'Univers of the Centre National de la Recherche Scientique of France, and the University of Hawaii.

Based on observations collected at the European Southern Observatory under ESO programmes 103.B-0568(A), 103.B-0568(B), 0104.B-0406(A) and 0104.B-0406(B).

Funding for SDSS-III has been provided by the Alfred P. Sloan Foundation, the Participating Institutions, the National Science Foundation, and the U.S. Department of Energy Office of Science. The SDSS-III web site is http://www.sdss3.org/.

SDSS-III is managed by the Astrophysical Research Consortium for the Participating Institutions of the SDSS-III Collaboration including the University of Arizona, the Brazilian Participation Group, Brookhaven National Laboratory, Carnegie Mellon University, University of Florida, the French Participation Group, the German Participation Group, Harvard University, the Instituto de Astrofisica de Canarias, the Michigan State/Notre Dame/JINA Participation Group, Johns Hopkins University, Lawrence Berkeley National Laboratory, Max Planck Institute for Astrophysics, Max Planck Institute for Extraterrestrial Physics, New Mexico State University, New York University, Ohio State University, Pennsylvania State University, University of Portsmouth, Princeton University, the Spanish Participation Group, University of Tokyo, University of Utah, Vanderbilt University, University of Virginia, University of Washington, and Yale University.

Guoshoujing Telescope (the Large Sky Area Multi-Object Fiber Spectroscopic Telescope LAMOST) is a National Major Scientific Project built by the Chinese Academy of Sciences. Funding for the project has been provided by the National Development and Reform Commission. LAMOST is operated and managed by the National Astronomical Observatories, Chinese Academy of Sciences.

\software{STREAMFINDER \citep{2018MNRAS.477.4063M}, Armadillo \citep{Sanderson2016}, IRAF \citep{1986SPIE..627..733T,1993ASPC...52..173T}}

\bibliography{Gaia_AllStreams}
\bibliographystyle{aasjournal}

\end{document}